\documentclass[10pt, conference, letterpaper]{IEEEtran}

\usepackage{cite}
\usepackage{amsmath,amssymb,amsfonts}
\usepackage{graphicx}
\usepackage{textcomp}
\usepackage{xcolor}
\usepackage{marvosym}

\usepackage{enumitem}
\usepackage{multirow}
\usepackage{subfigure}
\usepackage[linesnumbered,ruled,vlined]{algorithm2e}
\SetEndCharOfAlgoLine{}
\IEEEoverridecommandlockouts

\def\BibTeX{{\rm B\kern-.05em{\sc i\kern-.025em b}\kern-.08em
    T\kern-.1667em\lower.7ex\hbox{E}\kern-.125emX}}

\begin{document}

\title{ARTSN: Exact and Adaptive Self-triggered\\Traffic Scheduling for ARTS Networks}

\author{
    \IEEEauthorblockN{\ Ruide Cao$^{\mathbb{S}\mathbb{T}\mathbb{U}}$, Shuangping Zhan$^{\mathbb{P}}$, Jiashuo Lin$^{\mathbb{H}}$, Yan Liu$^{\mathbb{S}\mathbb{P}}$, Chenxi Ling$^{\mathbb{S}\mathbb{P}}$, Yi Wang$^{\mathbb{S}\mathbb{P}\mathbb{D}\text{\small\Letter}}$, Guoming Tang$^{\mathbb{T}\text{\small\Letter}}$}
    \IEEEauthorblockA{$^{\mathbb{S}}$Southern University of Science and Technology; $^{\mathbb{T}}$The Hong Kong University of Science and Technology (Guangzhou)}
    \IEEEauthorblockA{$^{\mathbb{U}}$University of California, Riverside; $^{\mathbb{P}}$Pengcheng Laboratory; $^{\mathbb{H}}$Hanshan Normal University;  $^{\mathbb{D}}$Heyuan DET}
    \thanks{\text{\small\Letter}\;Corresponding authors: wy@ieee.org, guomingtang@hkust-gz.edu.cn. This work is supported by the Guangdong Basic and Applied Basic Research Foundation (2024A0101010001), the Guangdong High-Level Talents Special Support Program (2021TX05X205), Pengcheng Laboratory The Major Key Project of PCL (PCL2025A01). This study was conducted by Ruide Cao during his internship at HKUST (GZ).}
}
\IEEEaftertitletext{\vspace{-3.5ex}}

\maketitle

\begin{abstract}
Autonomous real-time systems (ARTS), such as self-driving vehicles and robotic assembly lines, are increasingly deployed to improve efficiency, accuracy, and responsiveness with reduced human intervention. In ARTS networks, self-triggered (ST) traffic—initiated by internal decision-making rather than fixed schedules or external events—is becoming prevalent and plays a critical role in enabling timely autonomous actions. However, existing network schedulers do not adequately support ST traffic due to two inherent challenges: volatility, where bounded processing jitter leads to uncertain arrival times, and absence, where reserved network resources remain underutilized when ST traffic does not materialize. To address these challenges, we propose ARTSN, an ST-tailored scheduling paradigm built upon time-sensitive networking (TSN). ARTSN introduces two key techniques: (1) an exact offline scheduling method that leverages the inferable arrival information of ST traffic for precise time-slot reservation, and (2) an adaptive online slot-release mechanism that dynamically reclaims unused reservations when ST traffic is absent. Extensive experiments on both a TSN simulator and a real-world testbed show that ARTSN significantly improves schedulability, scalability, and efficiency over state-of-the-art methods while maintaining reliable transmission guarantees.
\end{abstract}


\section{Introduction}

Fueled by recent advances in embedded computing power and Deep Neural Networks (DNNs), end-to-end (E2E) control is evolving rapidly. A fundamental shift is \textbf{where} decisions are made. Modern systems increasingly make decisions internally on-device, rather than relying on human operators or cloud offloading. An example is Tesla’s Full Self-Driving (FSD) stack \cite{tesla_fsd}, where perception and driving decisions are computed onboard by a single DNN. Similar trends are emerging on smaller platforms as well: on an NVIDIA Jetson Nano, TensorRT-optimized DNNs can respond in tens of milliseconds (92.8\,ms for a 3D-CNN and 118.4\,ms for AlexNet) \cite{swaminathan2025benchmarking}. This intelligent onboard decision-making capability combines autonomy with real-time responsiveness, giving rise to Autonomous Real-Time Systems (ARTS). ARTS are substantially enhancing the efficiency, accuracy, and responsiveness of E2E control while simultaneously minimizing the need for human intervention. As a consequence, ARTS have attracted significant research interest~\cite{yang2018avoiding} and have been widely applied \cite{cao2024adaptive, de2023single, jover2023opportunities}.

\begin{figure}[tbp]
\centerline{\includegraphics[width=0.9\columnwidth]{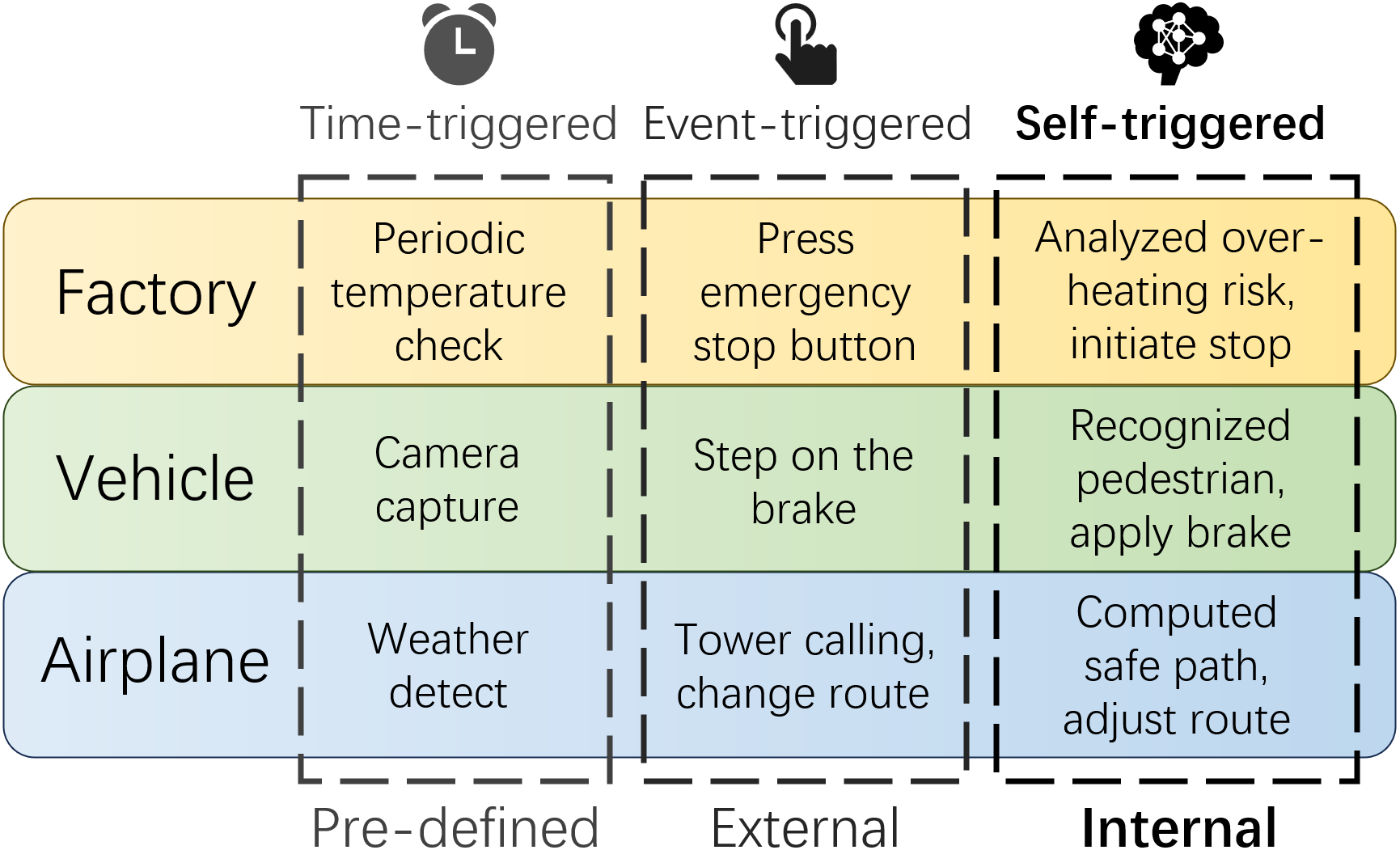}}
\caption{Real-world application examples for TT, ET, and ST control.}
\label{fig_scenarios}
\vspace{-0.4cm}
\end{figure}

Control tasks in ARTS are typically divided into three categories based on their triggering causes: time-triggered (TT), event-triggered (ET), and self-triggered (ST) \cite{heemels2012introduction}. TT tasks are scheduled at predetermined intervals for predictable and periodic operations. ET tasks are activated in response to specific external events or conditions, allowing for on-demand execution. In contrast, ST tasks are initiated proactively in response to internal state changes or criteria inherent to the system. Some real-world examples are listed in Fig.~\ref{fig_scenarios}. While TT tasks continue to predominate, ST tasks are increasingly pervasive \cite{yi2018dynamic}. The transition from pre-defined and external triggers to internal decision-making highlights the importance of ST tasks. In particular, the ST mechanism has distinct advantages in safety-critical scenarios, where waiting for a clock countdown or external decision-making signals before reacting is unacceptably slow~\cite{adimoolam2023safe}. From a network perspective, these control tasks correspond to distinct traffic types, each demanding strict latency and reliability guarantees.

To meet these stringent requirements, some ARTS networks in safety-critical domains rely on hardwiring~\cite{gent2020hosting}, where devices are connected through dedicated point-to-point physical links. Such a hardwiring approach is inherently inefficient and non-scalable, as adding new devices requires additional dedicated connections. In pursuit of reliable, efficient, and scalable traffic transmission, Time-Sensitive Networking (TSN) has emerged and attracted significant industrial interest due to its high flexibility and wide applicability \cite{zhang2024time, xue2026keepon}. As an Ethernet-based networking technology, TSN supports plug-and-play modular connectors. When applied in ARTS networks, TSN can naturally support TT traffic scheduling through its traffic shaping capabilities~\cite{8021qav,8021qbv,8021qch}. With recent research efforts, TSN can also be adopted for the ET traffic scheduling~\cite{zhao2022etsn}. Refer to \S~II-B for details.

When accommodating ST traffic with TSN, however, we are faced with the following threefold problem. 
\textbf{1) Failure to meet real-time scheduling requirements.} 
A straightforward approach is to leave ST traffic unprotected, \textit{i.e.}, transmit along with other best-effort (BE) traffic but with higher priority as in \cite{8021qav}. ST streams may still miss their deadlines due to the blocking of low-priority packets.
\textbf{2) Inefficient use of network bandwidth.} 
An alternative practical approach is to treat the ST traffic as ET and adopt the ET-enabled traffic scheduling mechanism. Nevertheless, treating ST traffic with inferable arrival times as completely random and unpredictable ET traffic may lead to overly conservative bandwidth reservations, thereby wasting substantial network capacity.
\textbf{3) Scalability limitation imposed by switch computing power.} 
Another approach is to leave the scheduling of ST traffic to an online scheduler~\cite{zhuge2024innet, jiang2024hebo}. Due to the limited computing power, however, this is only available for sporadic ST traffic. Deterministic transmission of bursty ST traffic is still not guaranteed (such as the GOOSE burst in substation communication networks~\cite{docquier2022relevance}).

To further bridge the gap between the flexibility required for autonomy and the determinism required for real-time, we are motivated to design an ST-tailored traffic scheduling paradigm. Specifically, by referring to the essential design principles of TSN, we present ARTSN in this work. In support of TT, ET, and ST tasks in ARTS, ARTSN aims to provide reliable, deterministic traffic scheduling, enabling critical traffic from these tasks to be protected with extremely low latency and hard real-time guarantees.

To the best of our knowledge, existing scheduling algorithms can only provide \textit{feasible} but \textit{inexact} (refer to Def. 3 and 4 in \S \ref{chap_exact_traffic}) scheduling schemes for ST traffic. This is the first work to include exact ST traffic scheduling on TSN. Our major contributions can be summarized as follows:
\begin{itemize}
\item We extend the flexibility boundaries of the deterministic network system model by introducing the self-triggering mechanism from control systems to the network traffic scheduling context.
\item We propose ARTSN, an ST-tailored TSN traffic scheduling paradigm that supports all 3 traffic types. An exact traffic scheduling technique that effectively leverages inferable arrival information from ST traffic is presented, along with an adaptive slot release mechanism.
\item We evaluate ARTSN in both simulations and a real-world testbed. Comprehensive experiments show that ARTSN significantly improves network schedulability, scalability, and efficiency while guaranteeing reliability.
\end{itemize}

\section{Related Work}

For decades, multiple technologies have been designed to guarantee the transmission of critical traffic over Ethernet's high speed and good scalability. This section provides a brief literature review on Ethernet-based deterministic networks.

\subsection{Reliable Transmission of Time-Triggered Traffic}

1) TSN, supported by Broadcom (BCM89571\cite{broadcom_BCM89571}), Xilinx, Intel, Cisco, NXP, etc. \cite{tsnmarket}; 2) Time-Triggered Ethernet (TTEthernet), selected by ESA and NASA for their Orion multi-purpose vehicle \cite{loveless2015ttethernet}; 3) Avionics Full-Duplex Switched Ethernet (AFDX), used in Airbus A350 and A380 airplanes \cite{soni2024hybrid}; 4) PROFINET \cite{profinet}; 5) EtherCAT \cite{ghandour2024real}. These technologies use time-division multiplexing to segregate critical traffic from non-critical traffic. Pre-defined TT critical traffic can get low-latency, high-reliability transmission \cite{garreau2023link}. Frame loss can be solved through replication rather than retransmission \cite{8021cb}.

Among these, TSN has attracted much industrial interest due to its wide applicability \cite{wang2023reinforcement, wang2024enabling, sasiain2024towards}. Recently, 
Yang~\textit{et al.}~\cite{yang2023caas} proposed an industrial control system architecture and a task-traffic joint scheduling algorithm in the TSN context, ensuring traffic determinism by assigning task computation to switches.

\subsection{Reliable Traffic Transmission in ARTS networks}

Pre-defined TT traffic lacks the flexibility required for environment-adaptive tasks and is insufficient to support ARTS \cite{cao2025sagkit}. In ARTS networks, traffic characteristics are not as fixed as in traditional real-time systems but rather more volatile and episodic \cite{li2019predicting}. Both online and offline scheduling solutions have been proposed to handle such non-fixed traffic.

For online solutions, D-TSN \cite{patti2022deadline} supports event-driven real-time traffic using earliest-deadline-first scheduling, enabling dynamic response to aperiodic arrivals. InNetScheduler \cite{zhuge2024innet} utilizes computing resources on switches to schedule latency-sensitive ET traffic, thereby increasing the amount of schedulable traffic without centralized coordination. Hebo \cite{jiang2024hebo} loosens arrival time constraints of periodic streams, making TSN applicable to volatile traffic with bounded jitter.

For offline solutions, E-TSN \cite{zhao2022etsn} enables ET traffic scheduling by allowing ET streams to preempt shareable TT streams. A "prudent slot reservation" technique is presented to compensate for all possible preemptions. Notably, E-TSN's extra reservation does not iterate until convergence: when the extended window admits more ET arrivals, further extension is needed, and the lack of this recursive calculation may lead to deadline violations under certain conditions (analyzed in \S \ref{chap_exact_traffic}). DeepScheduler \cite{he2023deepscheduler} applies deep reinforcement learning to enhance schedulability and efficiency for large-scale networks. GP-TSN \cite{li2025event} leverages causal inference to predict future ET streams by modeling event interactions using a spatio-temporal graph attention mechanism.

However, none of the above works address ST traffic natively. Unlike ET streams triggered by unpredictable external events, ST streams are initiated by internal system states and have deterministic triggering relationships with TT streams, enabling more precise scheduling. Yet ST traffic also poses unique challenges: bounded \textit{volatility} from processing jitter and potential \textit{absence} when not triggered. Existing ET-oriented approaches rely on probabilistic prediction for stochastic events, which cannot exploit the deterministic structure inherent in ST streams. This gap is the focus of our work.

\section{System Model}

\subsection{Network Model}

We model the network topology with a directed graph $G=(\mathcal{N}, \mathcal{L})$. All devices and switches are nodes in $\mathcal{N}$, and all links connecting two nodes are in $\mathcal{L} \subseteq \mathcal{N}^2$. Each pair of connected nodes $n_a$, $n_b$ is assumed to be connected by two edges $\left \langle n_a, n_b \right \rangle$ and $\left \langle n_b, n_a \right \rangle$, following the full-duplex context. Two attributes of the edges are considered: $\langle n_a, n_b \rangle.\text{pd}$ denotes the propagation delay of $\langle n_a, n_b \rangle$, and $\langle n_a, n_b \rangle.\text{tu}$ is its smallest time unit. We use $\{.\}$ to denote a set where the order is irrelevant and $\langle.\rangle$ to denote a sequence of items.

We consider all the network traffic as unicast streams, \textit{w.l.o.g.}, following \cite{steiner2010evaluation, craciunas2016scheduling}. The set of physically generated concrete traffic is referred to as \textit{actual streams}, denoted $R = \{r_1, r_2, \dots, r_n\}$. For the $i$th actual stream $r_i$, we represent its path by $P_i=\langle\langle n_a, n_b \rangle, \langle n_b, n_c \rangle, \dots, \langle n_y, n_z \rangle\rangle$, and its maximum allowed E2E latency by $D_i^R$. Based on how $r_i$ is triggered and whether it shares its time slot with ET streams, $r_i$ has a type $t_i \in \{\text{strict}, \text{share}, \text{event}, \text{self}\}$. While $t_i = \text{event}$ or self directly indicates $r_i$ is ET or ST, strict and share are two mutually exclusive types of TT streams. The priority order is: $\text{strict} > \text{event} = \text{self} > \text{share}$. Strict TT streams have the highest priority and require strict transmission time. ET and ST streams share the same priority level, as both require timely delivery upon arrival. Share TT streams have the lowest priority among critical traffic and can be preempted by ET or ST streams when they are present.

\begin{figure}[tbp]
\centerline{\includegraphics[width=0.92\columnwidth]{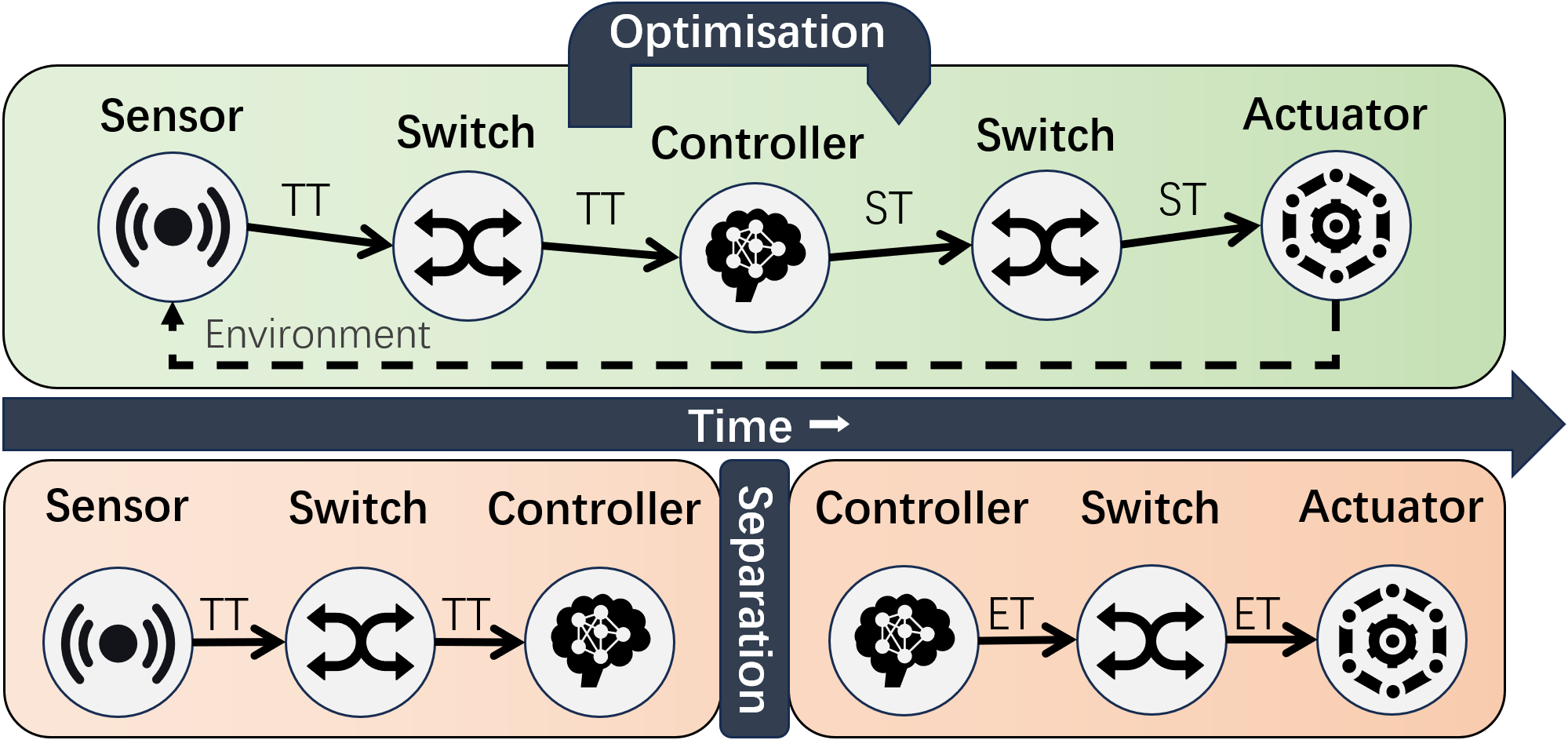}}
\caption{Transmission with (above) and without (below) the ST support. Joint considerations allow for the utilization of ST traffic arrival time information.}
\label{fig_arttsn}
\vspace{-0.3cm}
\end{figure}

We use $T_i$ to denote the period of TT streams and the minimum times between successive triggers for each ET and ST stream. There is a hyperperiod H of all actual streams $R$, where $H=lcm(T_1, T_2, \dots, T_n)$, which is the least common multiple of all $T_i$. Note that ST traffic is inherently bounded per hyperperiod, as ST streams are generated in response to TT streams, instead of being generated spontaneously by the controller. We then discuss every network scenario within its first hyperperiod, since later hyperperiods can be considered repetitions. A network scenario is defined as follows.

\textbf{Definition 1.} A \textit{network scenario} $\gamma=(G,R)$ consists of a network $G=(\mathcal{N},\mathcal{L})$ and the set of actual streams $R=\{ r_1,r_2,\dots,r_n \}$ in $G$.

In a network scenario, TT streams can be scheduled deterministically by setting proper offsets on the time-synchronized end nodes. For ET streams, however, the arrival times are independent of the system time. We denote the $j$th actual arrival time of ET stream $r_i$ as $A_{i,j}^R$. Furthermore, we assume each ST stream $r_k$ has a final trigger stream $tr_k$, which is the last TT stream involved in triggering $r_k$. Thus, all possible controller processing times of $tr_k$ can be represented as an interval $[C_k^{min}, C_k^{max}]$ and the processing jitter range is $\psi_k=C_k^{max}-C_k^{min}$. Based on these formulations, a trigger scenario is defined as follows. 

\textbf{Definition 2.} A \textit{trigger scenario} $\gamma^*=(G, R, A^R, C)$  is a specific instantiation of a network scenario $\gamma = (G, R)$. Arrival times $A^R = \langle A_{a,1}^R, A_{a,2}^R, \dots, A_{m,n}^R \rangle$ and process times $C=\langle C_{b,1}, C_{b,2}, \dots, C_{o,p}\rangle$ such that each arrival of each ET stream $r_i$ has an arrival time $A_{i,j}^R \in \left[0,H\right)$ and each trigger of each ST stream $r_k$ has a process time $C_{k,l} \in [C_k^{min}, C_k^{max}]$.

A traditional deterministic network scenario with full TT streams corresponds to a single trigger scenario because all streams are fixed, whereas a more flexible network scenario with ET or ST streams may typically involve multiple trigger scenarios. The number of trigger scenarios equals the number of combinations of all possible arrival and processing times. As time is discretely considered, the scenarios are finite.

Although jitter in the controller's processing time introduces some uncertainty, it remains feasible to infer the arrival of ST traffic. Fig. \ref{fig_arttsn} compares the traditional ET-enabled paradigm with our proposed ST-enabled paradigm. While the traditional paradigm independently reserves resources for each stream, we implicitly infer the arrival of each ST stream through the controller processing time and the triggering relationships between streams. Targeted reservations can be made according to those inferences.

\begin{figure}[tbp]
\centerline{\includegraphics[width=\columnwidth]{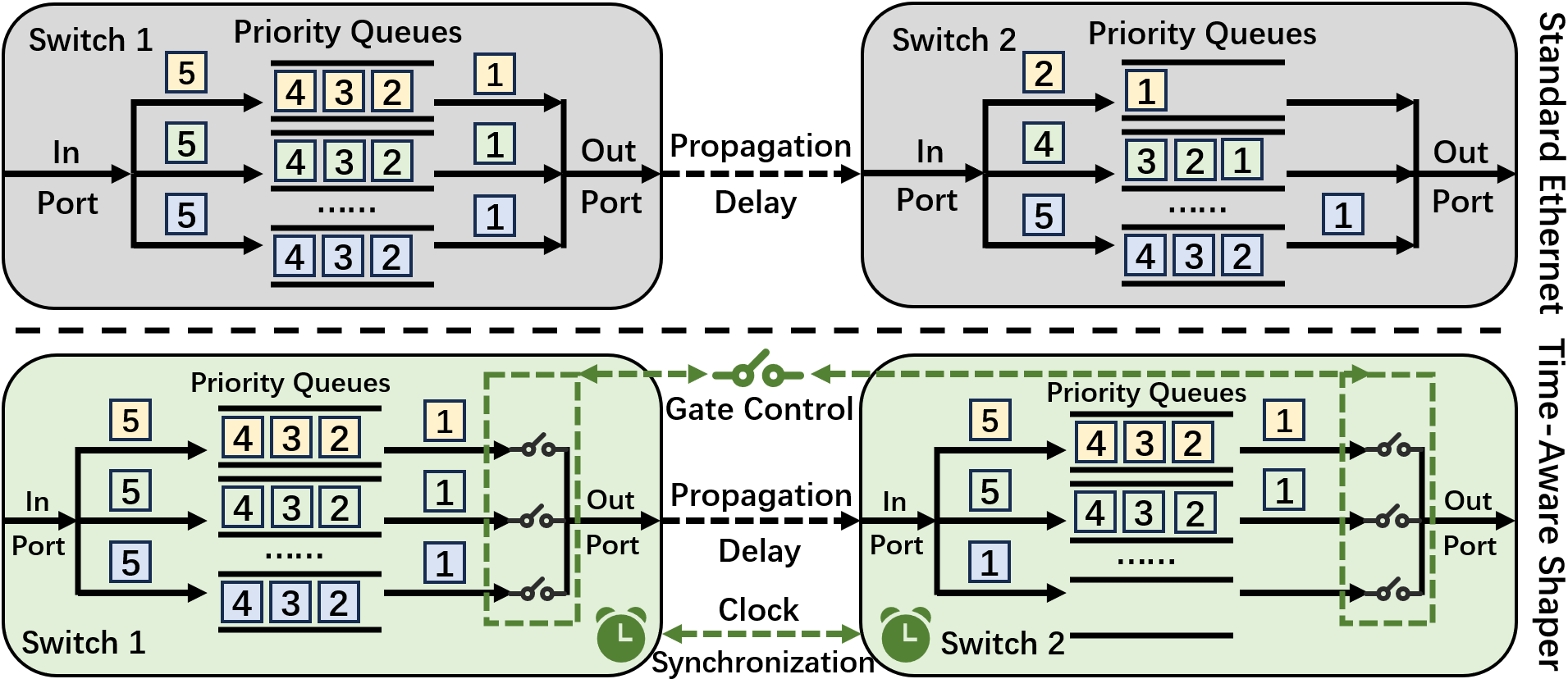}}
\caption{Standard Ethernet switch versus TAS switch.}
\label{fig_TAS}
\vspace{-0.25cm}
\end{figure}

\subsection{Switch Model}
\label{chap_switch_model}

The 802.1Q standard \cite{8021q} requires switches to support eight queues with different priorities. The 802.1AS standard \cite{8021as} enables global clock synchronization. Building on these standards, the 802.1Qbv \cite{8021qbv} ensures real-time performance through gate control. While queues spatially differentiate traffic, gating provides temporal differentiation, allowing for precise traffic scheduling. Fig. \ref{fig_TAS} illustrates the advancement of the Time-Aware Shaper (TAS) mechanism, which can achieve accurate traffic shaping via time-synchronized gating control.

As defined in the TSN framework \cite{8021qcc}, a Centralized Network Configuration (CNC) is deployed to calculate and distribute the Gate Control Lists (GCL). The TSN switches then open and close the egress port gates in synchronization with the local time and the GCL to avoid blocking caused by low-priority traffic. The gate opening time of each gate control entry in GCL is called a \textit{time slot}.

\subsection{Problem Definition}
\label{chap_formal_problem}
Based on the system models discussed above, we define the \textit{Deterministic Traffic Scheduling Problem}. The objective is to determine valid scheduling parameters and resource allocation for every actual stream in the network, ensuring deterministic guarantees for the arrival times of all traffic types.

\textbf{Given:} 
\begin{itemize}
    \item A network topology $G(\mathcal{N}, \mathcal{L})$ with link capabilities.
    \item A set of actual streams $R = \{r_1, r_2, \dots, r_n\}$, where each stream $r_i$ is characterized by its tuple $\langle T_i, D_i^R, P_i, \tau_i, t_i \rangle$. For ST streams, the trigger source $tr_k$ and processing jitter $\psi_k$ are also given.
\end{itemize}

\textbf{Find:}
\begin{itemize}
    \item \textbf{Time Slot Assignments $\Phi$:} A set of gate opening times for all frame instances that need to be scheduled. The detailed formulation of $\Phi$ will be presented in \S \ref{chap_exact_traffic}, where all actual streams are mapped to \textit{schedule streams}.
    \item \textbf{Queue Assignments $Q$:} The priority queue allocation for each stream, $Q = \{q_1, q_2, \dots, q_n\}$, where $q_i$ denotes the queue index for stream $r_i$.
\end{itemize}

\textbf{Subject to:}
\begin{enumerate}
    \item \textit{Hard Real-time Constraints:} All streams must complete transmission within their maximum allowed E2E latency $D_i^R$ and respect their period $T_i$.
    \item \textit{Network Constraints:} All frames must satisfy the physical constraints of links, including frame sequencing within each stream, propagation delays between adjacent links, and temporal isolation to prevent frame collisions at shared egress ports.
    \item \textit{Deterministic Constraints:} The derived schedule must be valid for \textit{all} possible trigger scenarios $\gamma^*$. For any actual arrival time within the processing jitter range $[C_k^{min}, C_k^{max}]$, the reserved time slots are sufficient to guarantee that no deadlines will be missed.
\end{enumerate}

\textbf{Objective:}
Find a tuple $\langle \Phi, Q \rangle$ such that all constraints are satisfied:
\begin{equation}
    \exists \langle \Phi, Q \rangle : \bigwedge_{c \in \mathcal{C}} c(\Phi, Q, G, R) = \text{True},
\end{equation}
where $\mathcal{C}$ represents the set of all aforementioned constraints. If such a solution exists, the network scenario is deemed \textit{schedulable}.

\begin{figure}[t]
\centerline{\includegraphics[width=0.95\columnwidth]{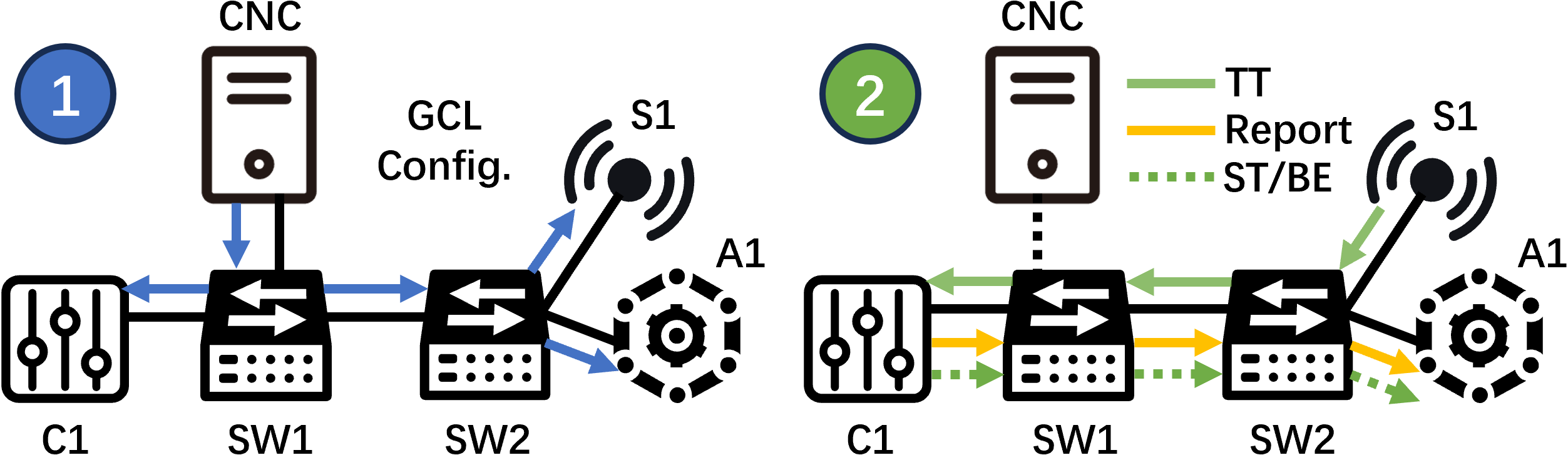}}
\caption{Offline (left) and online (right) scheduling phases of ARTSN.}
\label{fig_architecture}
\vspace{-0.3cm}
\end{figure}

\section{Design of ARTSN}
While \S \ref{chap_formal_problem} defines the general deterministic scheduling problem, existing solutions typically achieve only \textit{feasible}, but not \textit{exact}, scheduling schemes.
In this section, we first identify the design challenges and give an overview of ARTSN in \S~IV-A. We then introduce the concepts of \textit{feasible}, \textit{exact}, and present our design of exact traffic scheduling in \S~IV-B. We finally propose an adaptive slot release mechanism in \S~IV-C.

\subsection{Overview}

Prior knowledge of ST traffic can be used to improve both traffic schedulability and network efficiency. However, leveraging this knowledge in traffic scheduling requires addressing two design challenges: 
\begin{itemize}
    \item \textbf{Volatility.} Unlike TT streams with fixed arrival times, ST streams arrive within a bounded jitter window. This jitter arises from the controller's decision-making process, where different code paths under varying conditions lead to variable processing times. The scheduler must reserve slots covering all possible arrivals within this window—neither a single fixed slot (as for TT) nor slots spanning the entire period (as for ET) suffice.
    \item \textbf{Absence.} ST streams are conditionally triggered and may be absent in certain periods. For example, an autonomous vehicle only issues braking commands upon detecting obstacles—when the path is clear, no corresponding ST stream is generated. When absent, reserved slots still block other traffic, unnecessarily consuming bandwidth. This inefficiency is especially severe in topologies where ST streams converge, such as star and tree topologies.
\end{itemize}

To address the two challenges above and improve the performance of ARTS networks, we propose a two-phase traffic scheduling paradigm, ARTSN. Fig. \ref{fig_architecture} shows an overview of ARTSN in a sample network that includes a CNC, two switches, a sensor S1, a controller C1, and an actuator A1.
\textbf{1) In the offline scheduling phase}, we solve the volatility problem of ST traffic by modeling and constraining to make exact time slot reservations where ST streams are actually possible to arrive. These desired reservations are first calculated at the CNC, which then converts them into queue assignments and GCL configuration profiles. The profiles specify when each egress queue of each device can and cannot transmit. Once the calculations and conversions are complete, the CNC distributes them to all involved network devices. \textbf{2) In the online scheduling phase}, we solve the problem of idle reservation when ST traffic is absent by reporting its absence along its path in advance of the reserved time slots. Upon receiving a TT stream, the controller determines if a corresponding ST stream will be generated. This determination is made before the earliest possible transmission time for the ST stream, leaving a time gap. Thus, this time gap allows the controller to send a small BE report frame to release subsequent unnecessary reservations in non-triggering cases. If a release succeeds, more BE traffic can be transmitted, thereby improving network efficiency.

\subsection{Exact Traffic Scheduling}
\label{chap_exact_traffic}

To address volatility, exact traffic scheduling is designed in the offline scheduling phase. To clarify exactness, we first define a feasible traffic scheduling scheme as follows.

\textbf{Definition 3.} A traffic scheduling scheme is a \textit{feasible} traffic scheduling scheme iff \textbf{(i)} the scheme can be accomplished by deploying GCL at each egress port, and \textbf{(ii)} for all streams, there is no trigger scenario that may result in a deadline miss.

\begin{figure}[t]
\centerline{\includegraphics[width=0.9\columnwidth]{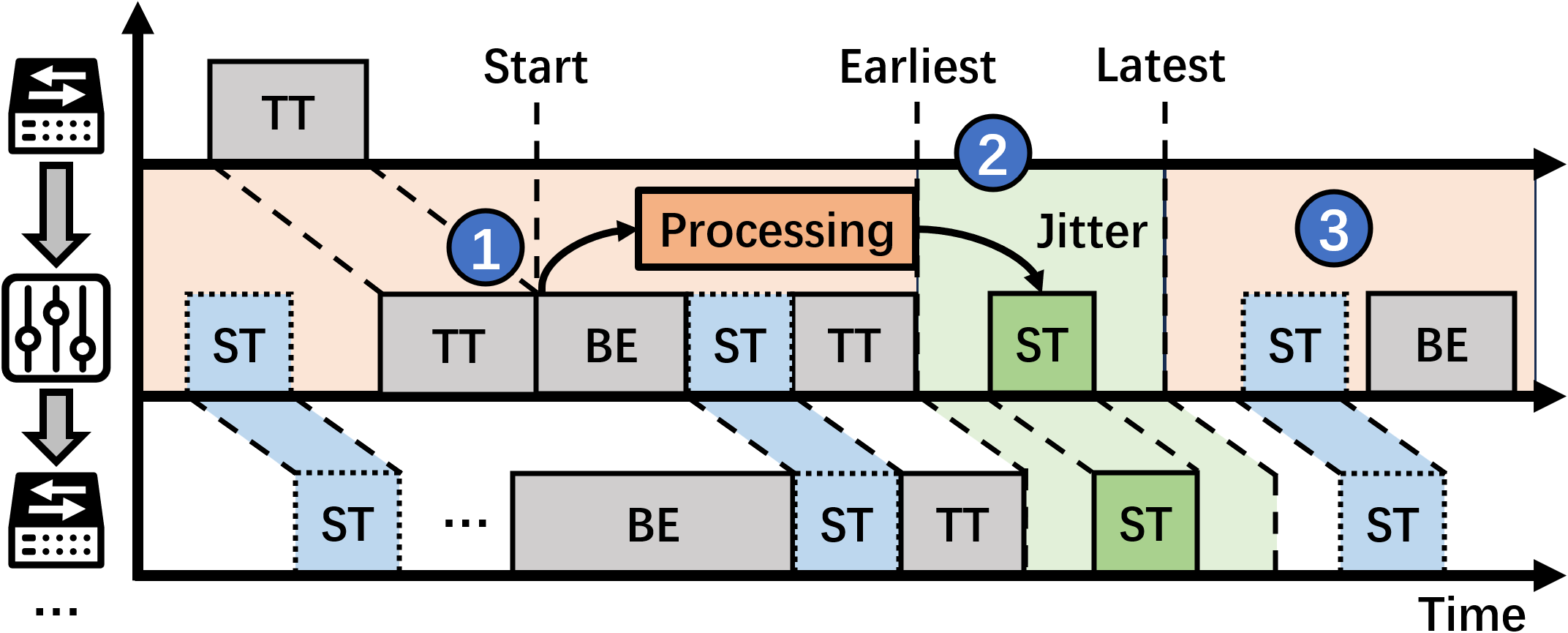}}
\caption{Three stages of an ST stream during one hyperperiod. The green time slot is reserved exactly, while the blue time slots are redundant reservations.}
\label{fig_exact}
\vspace{-0.25cm}
\end{figure}

A feasible traffic scheduling scheme ensures that all scheduled streams complete transmission before their deadlines. By treating each ST stream's minimum consecutive trigger interval as its period, all ST streams can be scheduled as TT streams. With such a transformation, several TT scheduling algorithms can schedule ST traffic by solving a Satisfiability Modulo Theories (SMT) or Integer Linear Programming (ILP) problem \cite{craciunas2016scheduling, yang2023caas, falk2018exploring}. 
These schemes are feasible when the maximum allowable latency of ST streams exceeds their minimum consecutive trigger interval, but their feasibility is not guaranteed when the latency requirement is more stringent. When the maximum allowable latency of an ST stream is less than its minimum consecutive trigger time, exponentially more time slots need to be reserved for the ST stream.

In addition to TT traffic, some approaches treat ST traffic as ET traffic to obtain feasible traffic scheduling schemes. As demonstrated in E-TSN \cite{zhao2022etsn}, the reliable transmission of ET traffic can be achieved by solving GCL constraints using SMT. E-TSN enables ET traffic to preempt some TT time slots, allowing ET traffic to arrive at any time without incurring deadline misses.

While feasible traffic scheduling schemes guarantee the delivery of the scheduled critical streams, they do not account for the impact on background traffic. In practice, the feasible traffic scheduling schemes identified by the above two approaches may result in a large number of idle reservation slots in the network. This underutilization occurs because many reserved time slots do not actually contribute to the transmission of ST streams while occupying the egress port, preventing other streams from being transmitted. We define an exact traffic scheduling scheme as follows.

\textbf{Definition 4.} A traffic scheduling scheme is an \textit{exact} traffic scheduling scheme iff \textbf{(i)} the scheme is \textit{feasible}, and \textbf{(ii)} for every reserved time slot, there exists at least one trigger scenario where its corresponding egress port is utilized during the slot, \textit{i.e.}, no reserved slot remains unused across all trigger scenarios.

Fig. \ref{fig_exact} illustrates the three stages within a hyperperiod from the controller's perspective. These three stages are separated by the earliest and latest possible completion times for the internal processing, respectively. In the first stage, the controller has not yet received or is processing the internal state information that may trigger the ST stream. In the second stage, the ST stream may arrive at any moment. In the third stage, the ST stream has either been generated or will not appear until the end of the hyperperiod. Therefore, reserving is valid only in the second stage.

To obtain an exact traffic scheduling scheme, we introduce \textit{schedule streams} $S = \{s_{1,1}, s_{1,2}, \dots, s_{n,m}\}$ as a collection of hypothetical streams. The frames of a schedule stream $s_{i,j}$ are $F_{s_{i,j}}^{\langle n_a,n_b \rangle}=\langle f_{s_{i,j},1}^{\langle n_a,n_b \rangle}, f_{s_{i,j},2}^{\langle n_a,n_b \rangle}, \dots, f_{s_{i,j},k}^{\langle n_a,n_b \rangle} \rangle$, and their start transmission times are $\Phi=\langle\phi_{s_{i,j},1}^{\langle n_a,n_b  \rangle}, \phi_{s_{i,j},2}^{\langle n_a,n_b  \rangle}, \dots, \phi_{s_{i,j},k}^{\langle n_a,n_b  \rangle}\rangle$. The transmission of $f_{s_{i,j},k}^{\langle n_a,n_b \rangle}$ takes $\tau_{s_{i,j},k}^{\langle n_a,n_b  \rangle}$ amount of time, and the maximum allowed E2E latency of $s_{i,j}$ is $D_{i,j}$. Since the transmission time is constant (given fixed bandwidth and maximum frame size), each slot can be represented by its start time only: the interval is simply $[\phi_{s_{i,j},k},\phi_{s_{i,j},k}+\tau_{s_{i,j},k}^{\langle n_a,n_b  \rangle}]$.

All the schedule streams are mapped from actual streams. Each TT actual stream is mapped to a single schedule stream, regardless of whether it shares time slots. Each ET stream maps to $N$ schedule streams, while each ST stream is mapped to $M$ schedule streams. Such one-to-many mappings are intended to characterize the volatility of their arrivals for scheduling. By discretizing all arrival possibilities of an ET actual stream $r_i$ into $N$ cases, the arrival time of its $j$th mapped schedule stream $s_{i,j}$ can be written as $A_{i,j} = T_i \times \frac{j}{N}$. For ST streams,
\begin{equation}
A_{i,j} = \phi_{s_{tr_i,1},last}^{P_{tr_i}[last]} + \tau_{s_{tr_i,1},last}^{P_{tr_i}[last]} + C_i^{min} + \psi_i \times (\frac{j}{M}),
\label{ST_arrive}
\end{equation}
where $\psi_i=C_i^{max}-C_i^{min}$ indicates the processing time interval of ST stream $s_i$, and $tr_i$ is the TT stream that may ultimately trigger $s_i$.

\begin{algorithm}[t]
\caption{Exact traffic scheduling of ARTSN}
\label{alg:solve_gcl}
\KwIn{Network $G$, actual streams $R$}
\KwOut{Start times $\Phi$, Queue assignment $Q$}
$S \gets \{\}$;\;
\ForAll{$r_i \in R$}{
    \If{$t_i \in \{\mathrm{strict, share}\}$}{
        Initialize $s_{i,1}$ according to $r_i$;\;
        \If{$t_i = \mathrm{share}$}{
            \Repeat{no $\tau$ changes}{
                \ForAll{$r_j \in R$ where $(t_j = \mathrm{event} \wedge \langle n_a, n_b \rangle \in P_i \cap P_j)$}{
                    $\tau_{et} \gets \tau_{j,1}^{\langle n_a, n_b \rangle} \times \left\lceil \frac{\tau_{i,1}^{\langle n_a, n_b \rangle}}{T_j} \right\rceil$;\;
                    $\tau_{i,1}^{\langle n_a, n_b \rangle} \gets \tau_{i,1}^{\langle n_a, n_b \rangle,\mathrm{orig}} + \tau_{et}$;\;
                }
            }
        }
    }
    \ElseIf{$t_i = \mathrm{event}$}{
        Initialize $\langle s_{i,1}, s_{i,2}, \dots, s_{i,N} \rangle$ with $D_{i,j}=D_i^R - \frac{T_i}{N}, A_{i,j}= T_i \times \frac{j}{N}$;\;
    }
    \ElseIf{$t_i = \mathrm{self}$}{
        Initialize $\langle s_{i,1}, s_{i,2}, \dots, s_{i,M} \rangle$ with $D_{i,j}=D_i^R - \frac{\psi_i}{M}, A_{i,j}$ set by Eq. (\ref{ST_arrive});\;
    }
    Add all generated $s_{i,j}$ to $S$;\;
}
Schedule constrained $S$ using SMT solver for {$\Phi$, $Q$};\;
\Return{$\Phi$, $Q$}
\end{algorithm}

With the above notations, an exact traffic scheduling scheme can be obtained by Alg. \ref{alg:solve_gcl}. Generally, we first convert all actual streams into schedule streams respectively, then formulate the schedule streams scheduling problem with a set of constraints, and finally solve the problem using an SMT solver.

The inputs of Alg. \ref{alg:solve_gcl} are actual streams $R$ and the network $G$. An empty schedule stream set $S$ is first initialized. During the traversal of each actual stream, the stream's type determines the rules according to which it will be converted.
\begin{enumerate}
\item A strict stream retains the original values of all its attributes to ensure that it gets strictly scheduled.

\item In contrast, the transmission time for each share stream is extended based on its per-hop ET demand. The extension must be computed \textit{recursively}: when a share stream's transmission time $\tau_{i,1}^{\langle n_a, n_b \rangle}$ increases, the number of possible ET preemptions $\lceil \tau_{i,1}^{\langle n_a, n_b \rangle} / T_j \rceil$ may also increase, requiring further extension. The algorithm iterates until a fixed point is reached, at which point no transmission time changes. This recursive extension, overlooked in prior work such as E-TSN~\cite{zhao2022etsn}, ensures that share streams have sufficient reservation to accommodate all possible ET preemptions without missing their deadlines.

\item Random arrivals of every event stream are uniformly divided into N parts for schedule consideration. The $j$-th schedule stream assumes arrival at time $A_{i,j} = T_i \times \frac{j}{N}$, with deadline $D_{i,j} = D_i^R - \frac{T_i}{N}$.

\item Similarly, the randomness of the occurrence time of each self stream is considered by means of M uniform possible intervals. The deadline for each possibility also subtracts the interval length. Assigning larger values for N and M will lead to more frequent redundant reservations, enhancing real-time performance but eroding overall network utilization.
\end{enumerate}

We formalize the criteria for exact scheduling schemes (there can be more than one) as the following constraints:

\textbf{Offset Range Constraint.} Each frame needs to be transmitted within its period. Therefore, a frame cannot start earlier than the period itself nor later than the latest possible start time to complete transmission within the period.
\begin{equation}
\begin{aligned}
\forall s_{i,j} & \in S, \forall \langle n_a, n_b \rangle \in P_i, \forall f_{s_{i,j},k}^{\langle n_a, n_b \rangle} \in F_{s_{i,j}}^{\langle n_a, n_b \rangle}:\\
& 0 \le \phi_{s_{i,j},k}^{\langle n_a, n_b \rangle} \le T_i - \tau_{s_{i,j},k}^{\langle n_a, n_b \rangle}.
\end{aligned}
\end{equation}

\textbf{Arrival Time Constraint.} The transmission of ET and ST streams can only start after arriving at the first node.
\begin{equation}
\forall s_{i,j} \in S,\ (t_i \in \{\text{strict}, \text{share}\}) \vee (\phi_{s_{i,j},1}^{P_i[1]}\ge A_{i,j}).
\end{equation}

\textbf{Frame Sequence Constraint.} Even when multiple streams mix at the same path, the sending sequence of their frames must maintain the original sequence within each stream.
\begin{equation}
\begin{aligned}
\forall s_{i,j} & \in S, \forall \langle n_a, n_b \rangle \in P_i, \forall f_{s_{i,j},k}^{\langle n_a, n_b \rangle} \in F_{s_{i,j}}^{\langle n_a, n_b \rangle}:\\
\mathrm{if\ } & f_{s_{i,j},k}^{\langle n_a, n_b \rangle} \mathrm{is \ not \ } f_{s_{i,j},last}^{\langle n_a, n_b \rangle}:\\
& \phi_{s_{i,j},k}^{\langle n_a, n_b \rangle} + \tau_{s_{i,j},k}^{\langle n_a, n_b \rangle} \le \phi_{s_{i,j},k+1}^{\langle n_a, n_b \rangle}.
\end{aligned}
\end{equation}

\textbf{End-to-End Latency Constraint.} The difference between a frame's arrival time and its complete time must not exceed the maximum allowed E2E latency.
\begin{equation}
\begin{aligned}
\forall & s_{i,j} \in S,\\
& (t_i \in \{\text{event}, \text{self}\} \wedge \phi_{s_{i,j},last}^{P_i[last]} - A_{i,j} \le D_{i,j})\ \vee \\
& (t_i \in \{\text{strict}, \text{share}\} \wedge \phi_{s_{i,j},last}^{P_i[last]} - \phi_{s_{i,j},1}^{P_i[first]} \le D_{i,j}).
\end{aligned}
\end{equation}

\textbf{Frame Overlap Constraint.} When two frames are going to be sent from the same egress port of the same switch, they can overlap in only two cases: 1) These two frames are derived from the same ET or ST stream; 2) One of them is ET, and the other is shareable TT. Other than the two cases, the sending time windows of the two frames must not overlap.
\begin{equation}
\begin{aligned}
\forall & \langle n_a,n_b \rangle \in \mathcal{L},\ \forall F_{s_{i,j}}^{\langle n_a,n_b \rangle}, F_{s_{k,l}}^{\langle n_a,n_b \rangle}, i \ne k,\\
\{&t_i\}\cup\{t_k\} \ne \{\text{event},\text{share}\}:\\
\forall & f_{s_{i,j},g}^{\langle n_a,n_b  \rangle}\in F_{s_{i,j}}^{\langle n_a,n_b \rangle}, f_{s_{k,l},h}^{\langle n_a,n_b  \rangle}\in F_{s_{k,l}}^{\langle n_a,n_b \rangle},\\
\forall & x \in \{1,2,\dots,lcm(T_i,T_k)/T_i\},\\
\forall & y \in \{1,2,\dots,lcm(T_i,T_k)/T_k\}:\\
& (\phi_{s_{i,j},g}^{\langle n_a,n_b \rangle} + x \times T_i \ge \phi_{s_{k,l},h}^{\langle n_a,n_b \rangle} + y \times T_k + \tau_{s_{k,l},h}^{\langle n_a,n_b  \rangle}) \ \vee \\
& (\phi_{s_{k,l},h}^{\langle n_a,n_b \rangle} + y \times T_k \ge \phi_{s_{i,j},g}^{\langle n_a,n_b \rangle} + x \times T_i + \tau_{s_{i,j},g}^{\langle n_a,n_b  \rangle}).
\end{aligned}
\end{equation}

\textbf{Queue Assignment Constraint.} The switches achieve spatial isolation of streams by queues. We reserve one dedicated queue for ET streams and one for ST streams. TT streams that do not share time slots with ETs use the queues from $NSH_L$ to $NSH_H$, and TT streams that perform time slot sharing with ETs use the queues from $SH_L$ to $SH_H$.
\begin{equation}
\begin{aligned}
\forall s_{i,j} & \in S: &\\
& (t_i = \text{strict} \ \wedge \ NSH_L \le q_i \le NSH_H)\ \vee\\
& (t_i = \text{self} \ \wedge \ q_i = ST)\ \vee\\
& (t_i = \text{event} \ \wedge \ q_i = ET)\ \vee\\
& (t_i = \text{share} \ \wedge \ SH_L \le q_i \le SH_H).
\end{aligned}
\end{equation}

In addition to the above constraints, the frame isolation constraint (Appendix \ref{appendix_frame_iso}) and the adjacent link constraint (Appendix \ref{appendix_adjacent_link}) are also needed in line with \cite{craciunas2016scheduling, zhao2022etsn}. With all these constraints, an exact traffic scheduling scheme can be obtained with the Z3 solver \cite{z3prover}. According to the obtained scheme, Alg. \ref{alg:solve_gcl} eventually outputs the times $\Phi$ when each frame starts to be transmitted in every hyperperiod and the queue assignment $Q=\{q_1, q_2, \dots, q_{n}\}$ of all actual streams.

\subsection{Adaptive Slot Release}

To address the resource waste caused by the absence of ST streams, we design an adaptive slot release mechanism (Fig. \ref{fig_adaptive}) with three phases: a \textit{processing phase}, a \textit{reporting phase}, and a \textit{releasing phase}. The structure of this section is organized accordingly.

\subsubsection{Processing Phase}
The controller extracts the payloads of the TT frames and performs calculations to make decisions. Distinguished from the time when switches and NICs process the data frame headers, this processing is handled by applications running on top of the controller. After processing, whether the corresponding ST stream needs to be triggered can be determined from the next system action indicated by the controller's processing result.

\subsubsection{Reporting Phase}
\begin{figure}[tbp]
\centerline{\includegraphics[width=0.9\columnwidth]{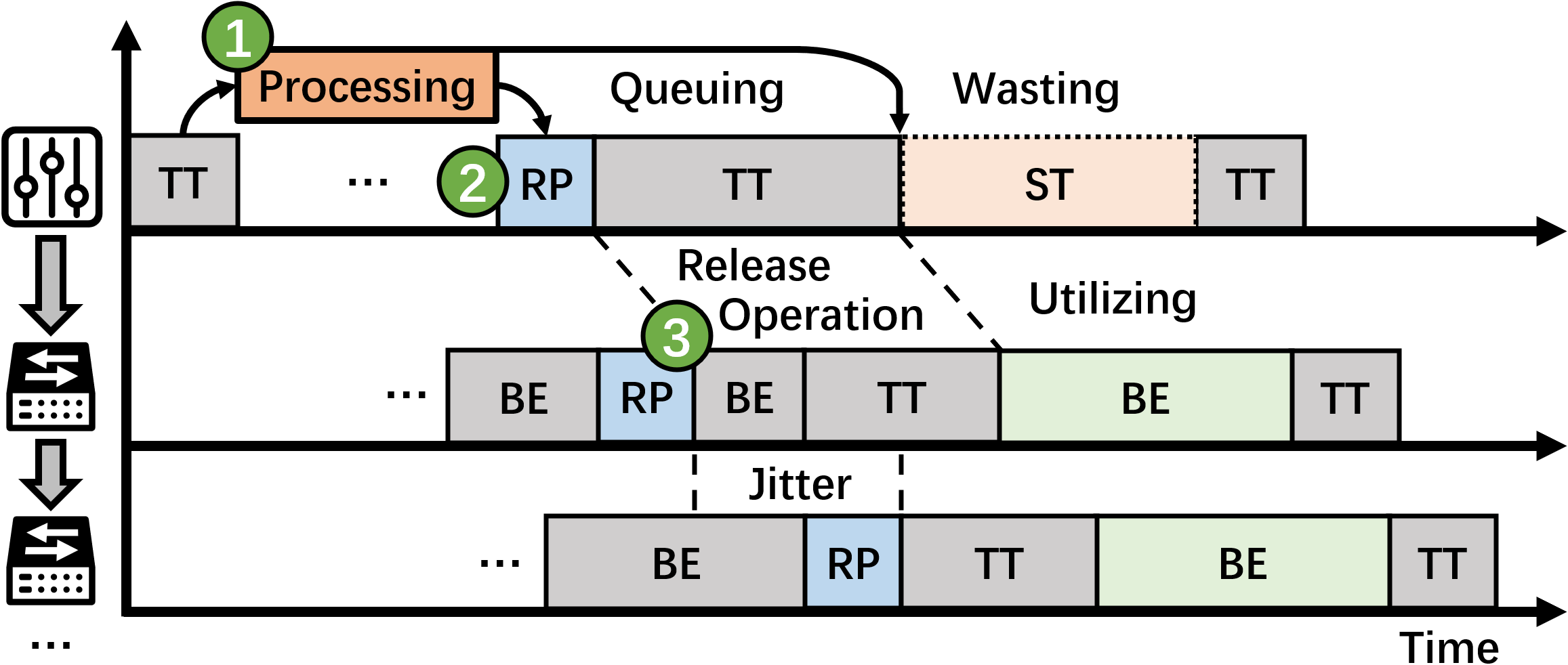}}
\caption{Three phases of the adaptive slot release mechanism.}
\label{fig_adaptive}
\vspace{-0.3cm}
\end{figure}

As soon as the controller becomes aware that it no longer needs to produce the ST stream, it can attempt to report its absence to downstream switches by transmitting a small report frame. This frame (marked as RP in Fig. \ref{fig_adaptive}) requires only a unique identifier, along with the ID of the reported ST stream and the hyperperiod. The controller sends this frame as BE traffic. Since this RP frame is sent from the same source but prior to the ST stream, it is expected to arrive at each hop ahead of the reserved time slots of the ST stream, although this is not guaranteed.

\subsubsection{Releasing Phase}
\begin{figure}[tbp]
    \centering
    \begin{minipage}[t]{0.23\textwidth}
        \centering
        \includegraphics[scale=0.108]{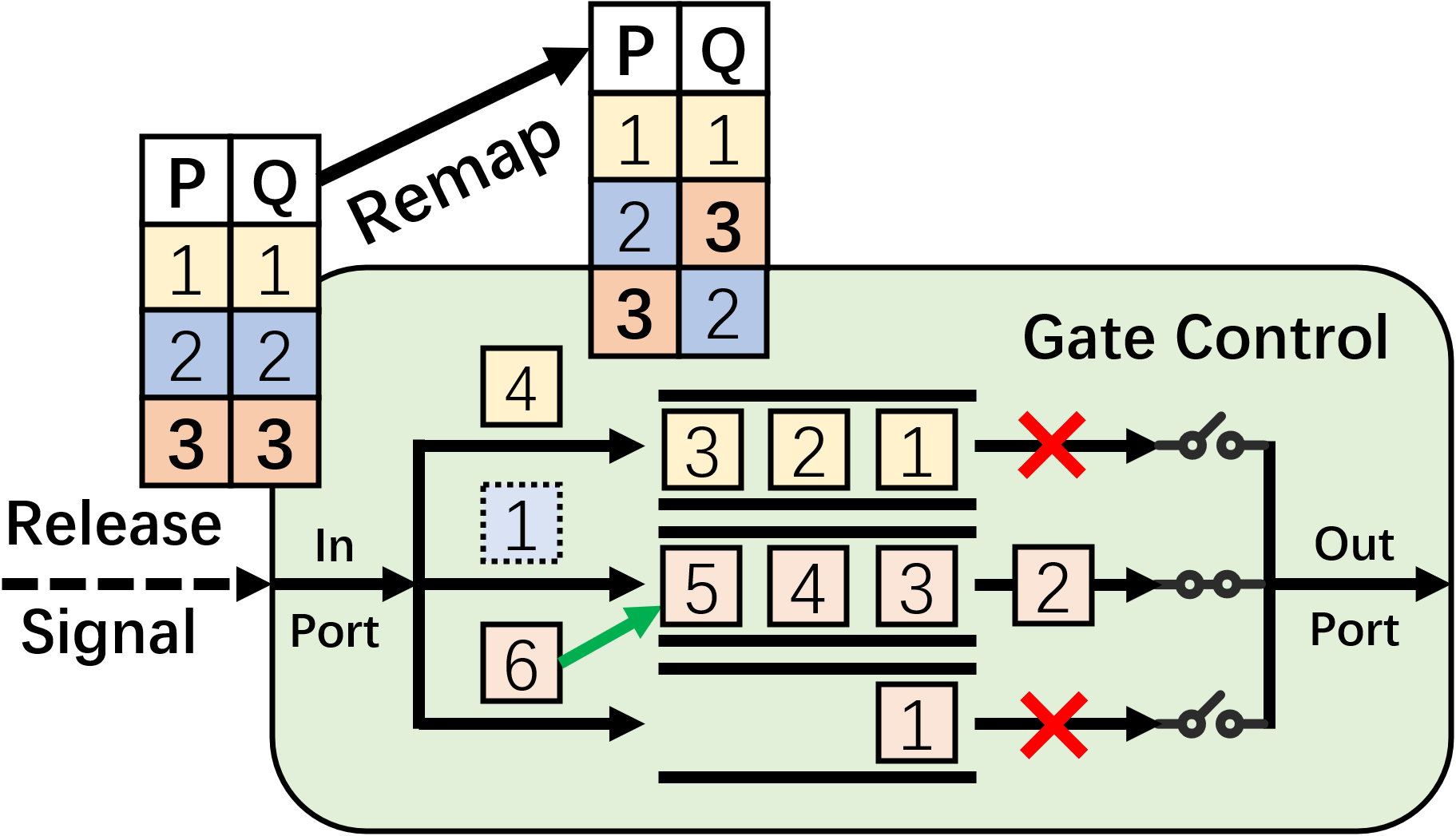}
        \caption{Mapping modification.}
        \label{fig_remap}
    \end{minipage}
    \begin{minipage}[t]{0.25\textwidth}
        \centering
        \includegraphics[scale=0.10]{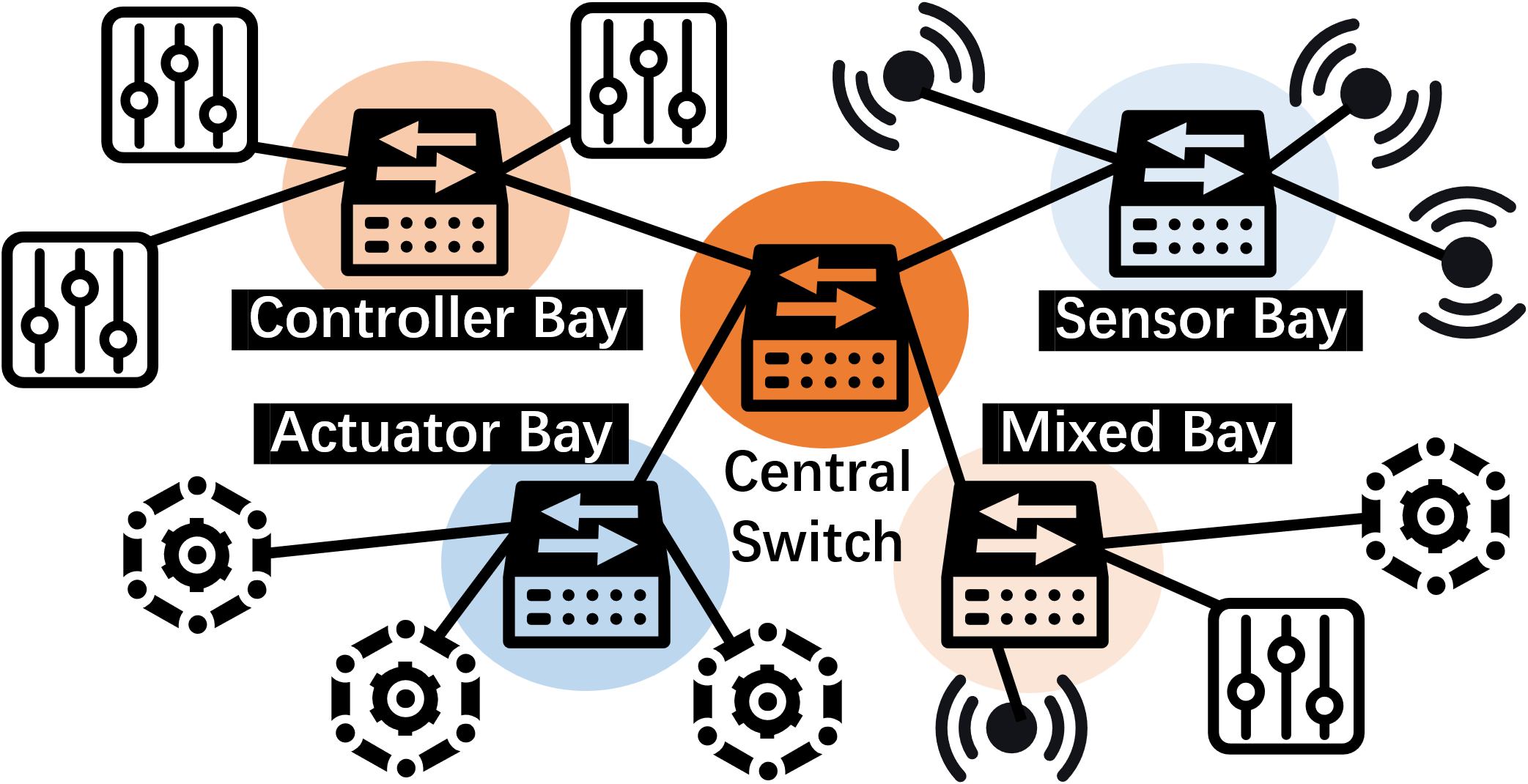}
        \caption{Concentrated waste.}
        \label{fig_star_st}
    \end{minipage}
\vspace{-0.3cm}
\end{figure}

Once a downstream switch receives a frame reporting the absence of a particular ST stream and specifying the current hyperperiod, the mapping relationship between frame priority and switch queues is modified during that absence. By this means, the switches enable BE traffic to enter the ST queue for transmission during the absence of ST streams, as Fig. \ref{fig_remap} shows. While modifying the GCL to open the BE queue gate does not take effect in real-time, this mapping modification operation can achieve the same effect in \textbf{microseconds} with hardware support.


This mechanism can adapt to the triggering conditions of ST streams to dynamically regulate network transmission, thereby improving efficiency. However, it requires consideration of network resource overhead (for transmitting RP frames) and hardware support from the switches. For these reasons, the mechanism should be activated selectively on certain switches at certain times. For example, as shown in Fig. \ref{fig_star_st}, the worst waste occurs at the central switch, where enabling the mechanism yields the greatest benefit. We call it low-intrusive because other switches do not require any changes.

Given the minimum Ethernet frame size of 64 bytes, we designed the RP frame to be 70 bytes. Since the RP frames are transmitted as BE traffic without real-time guarantees, release failures may occur. Thus, the mechanism should be activated only when ST streams are relatively heavily loaded and volatile, as the payback and success rate of releasing are higher. The mathematical expectation of the release gain for each ST stream applied should be at least 70 bytes.

\section{Evaluation}

This section starts with environment setup and baselines (\S \ref{section_setup} and \S \ref{section_baselines}), followed by comprehensive performance comparisons in terms of \textbf{schedulability} (\S \ref{section_schedulability}), \textbf{scalability} (\S \ref{section_scalability}), \textbf{efficiency} (\S \ref{section_efficiency}), and concludes with the \textbf{reliability} validation (\S \ref{section_validation}). Six key metrics are included: 1) number of schedulable scenarios, 2) memory consumption, 3) runtime, 4) theoretical utilization, 5) E2E latency, and 6) E2E jitter.

Our experiments lead to the following key findings:
\begin{itemize}
\item ARTSN's schedulability significantly outperforms other approaches in ST-contained scenarios, enabling effective scheduling even when all other approaches fail.
\item ARTSN consumes the least memory and the shortest runtime under increasing network loads, demonstrating the best scalability with the slowest overhead growth.
\item ARTSN is exceptionally efficient in ST-contained scenarios. It exactly reserves all valid time slots and may even achieve zero waste through dynamic release.
\item Verified by hardware in practice, ARTSN can guarantee critical transmissions under background traffic disturbances. In our 5-hop testbed, it achieved a 47.18 µs worst-case latency and a 1.01 µs maximum jitter over 5 minutes.
\end{itemize}

\subsection{Environment Setup}
\label{section_setup}

\begin{table}[b]
\vspace{-0.3cm}
\caption{Stream pattern of evaluation}
\centering
\resizebox{\columnwidth}{!}{
\begin{tabular}{|c|ccccccccc|}
\hline
\textbf{Number}       & \multicolumn{1}{c|}{5} & \multicolumn{1}{c|}{10} & \multicolumn{1}{c|}{5}  & \multicolumn{1}{c|}{5} & \multicolumn{1}{c|}{10} & \multicolumn{1}{c|}{5}  & \multicolumn{1}{c|}{1}    & \multicolumn{1}{c|}{1}    & 1    \\ \hline
\textbf{Type}         & \multicolumn{3}{c|}{share}                                                 & \multicolumn{3}{c|}{strict}                                                & \multicolumn{3}{c|}{self}                                    \\ \hline
\textbf{Size (bytes)} & \multicolumn{7}{c|}{500 * $U$}                                                                                                                                                      & \multicolumn{2}{c|}{1500 * $U$}  \\ \hline
\textbf{Period (ms)}  & \multicolumn{1}{c|}{5} & \multicolumn{1}{c|}{10} & \multicolumn{1}{c|}{20} & \multicolumn{1}{c|}{5} & \multicolumn{1}{c|}{10} & \multicolumn{1}{c|}{20} & \multicolumn{1}{c|}{10}   & \multicolumn{1}{c|}{10}   & 10   \\ \hline
\textbf{Probability}      & \multicolumn{6}{c|}{100\%}                                                                                                                              & \multicolumn{1}{c|}{50\%} & \multicolumn{1}{c|}{70\%} & 80\% \\ \hline
\end{tabular}
}
\label{table_flow}
\end{table}

\begin{figure}[t]
    \centering
    \subfigure{\includegraphics[scale=0.103]{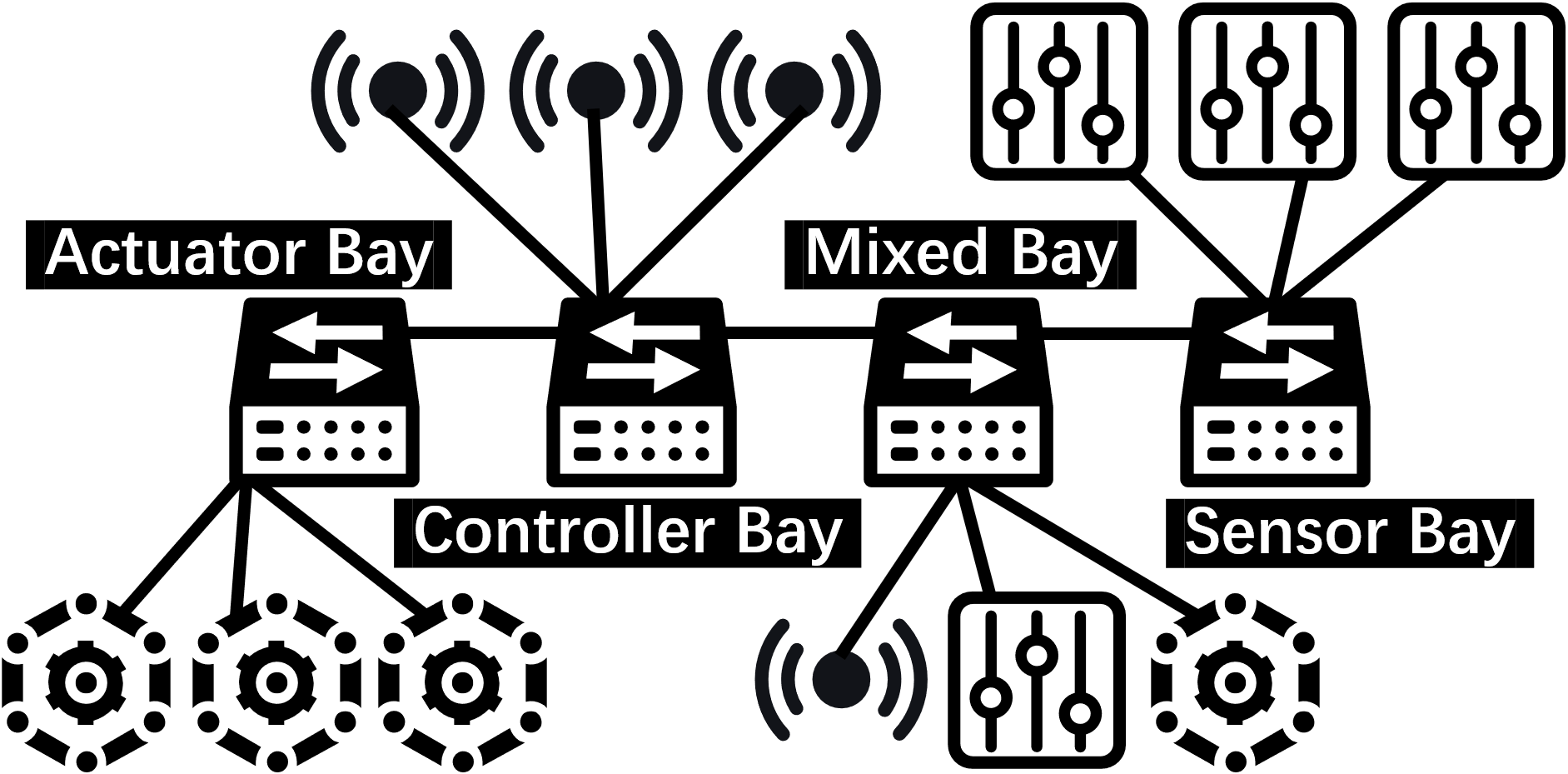}}
    \subfigure{\includegraphics[scale=0.117]{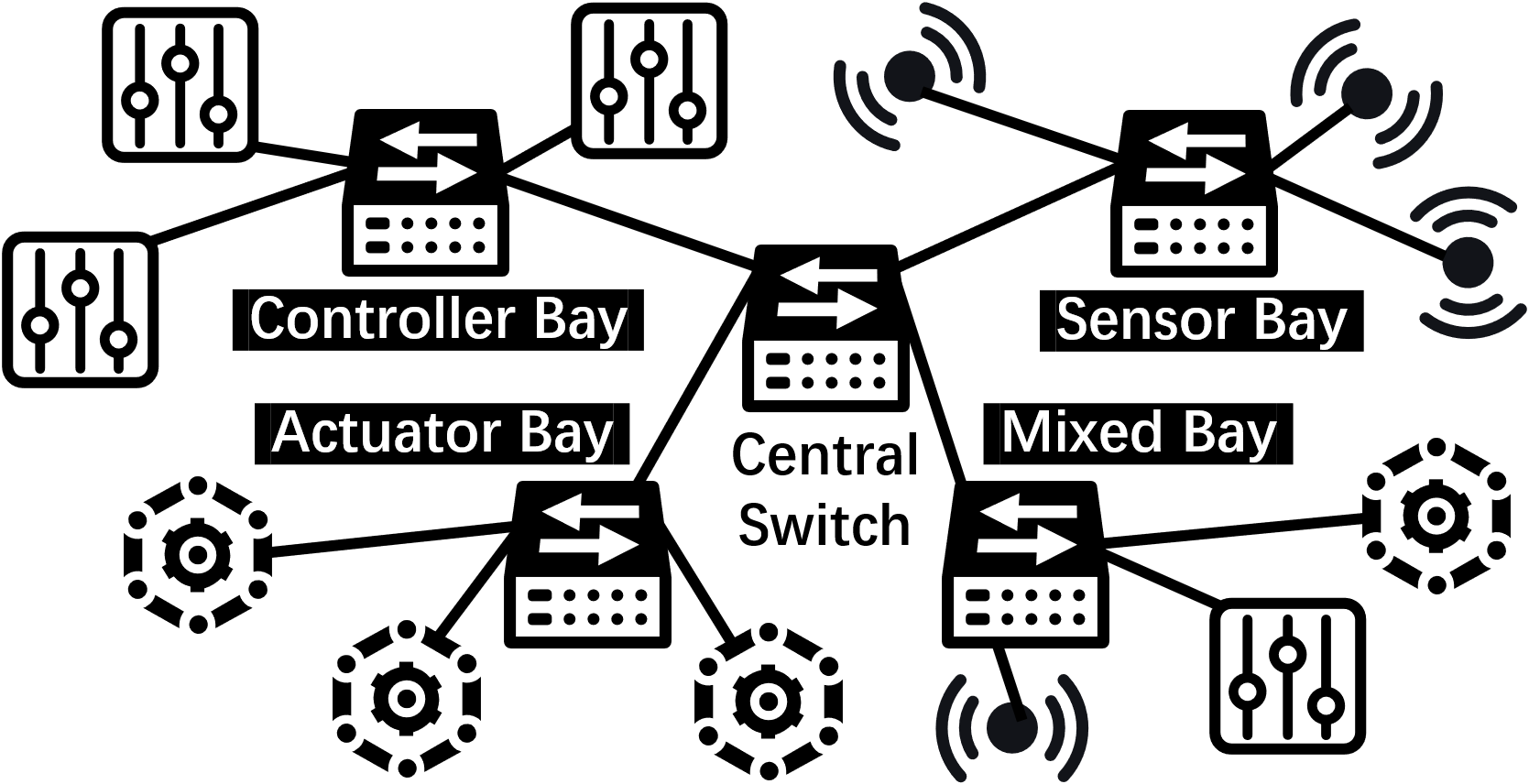}}
    \caption{Line topology (left) and star topology (right) with 4 bays.}
    \label{fig_topos}
\vspace{-0.25cm}
\end{figure}

Referring to the case studies on substation communication networks \cite{docquier2022relevance} and in-vehicle networks \cite{peng2023survey}, we conduct experiments with two representative topologies, as in Fig. \ref{fig_topos}. Each local switch and all its connected end-devices form a \textit{bay}. There are 4 bays in each topology. The sensor, controller, and actuator bays are connected to three of their corresponding end-devices, while the mixed bay is connected to one of each kind. The difference between the two topologies is that, in the \textit {line} topology, the local switches are directly connected to a single line, whereas in the \textit{star} topology, they are collectively connected to a central switch.

According to the IEC 61850 requirements for GOOSE traffic on real-time substation control \cite{docquier2020iec} and the event tracking performance of embedded SCADA systems \cite{docquier2022relevance}, the stream workload is set as in Table \ref{table_flow}. All links have 1 Gbps bandwidth. A total of 43 streams are in our test networks: 40 are TT streams, and 3 are ST streams. To accommodate more practical application scenarios, the 3 ST streams are set to different sizes or trigger chances. We use a utilization factor $U$ to scale the streams from 500 bytes (1 Ethernet frame) to 150,000 bytes (100 Ethernet frames) to achieve different network utilization rates. Each stream has an E2E deadline equal to its period, and the hyperperiod is 20 ms. The ST streams share a 1 ms processing-time jitter, meaning a 1 ms interval between the earliest and latest completion times for the internal ST-triggering processing.

\begin{figure}[t]
\centerline{\includegraphics[width=0.8\columnwidth]{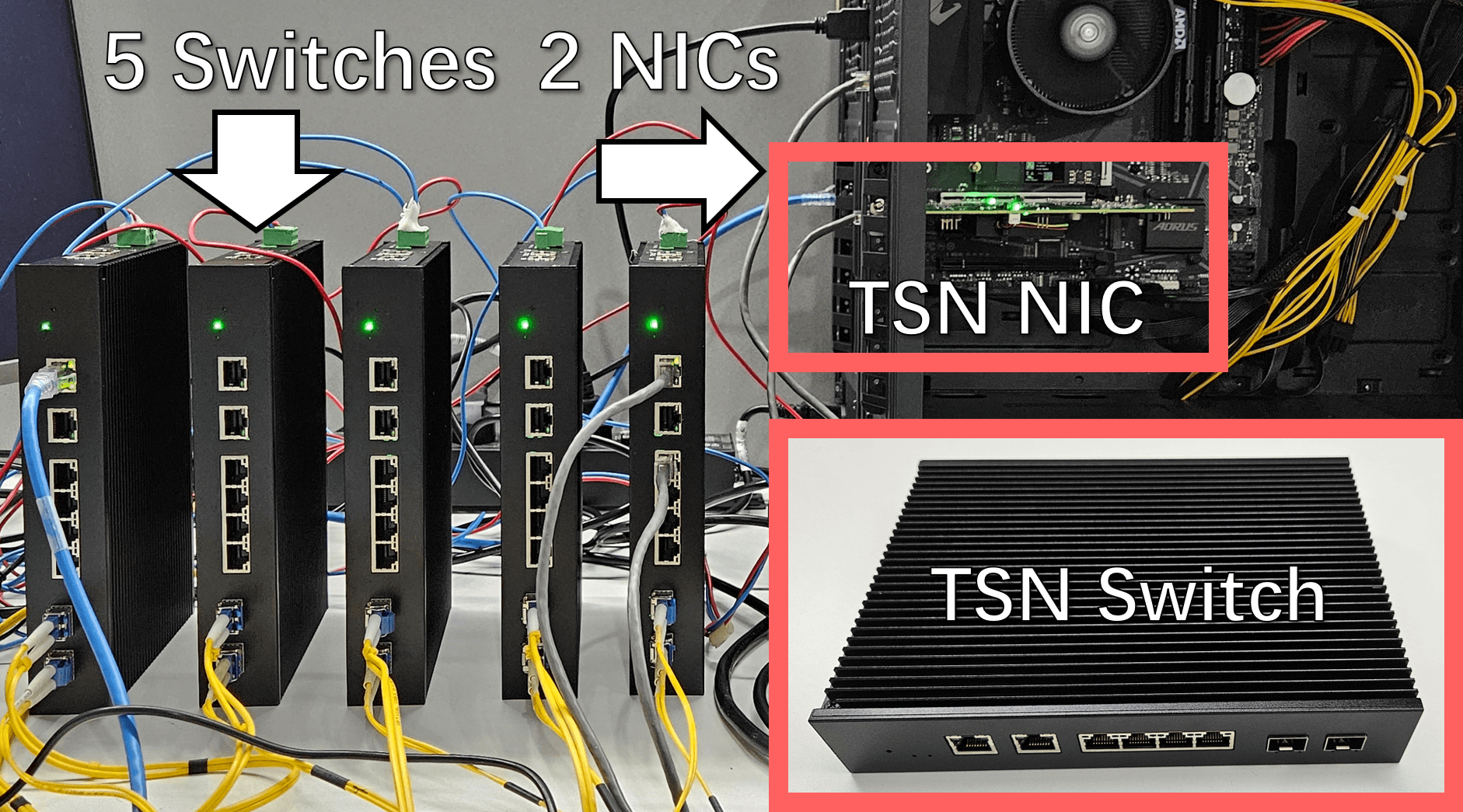}}
\caption{Testbed with 5 TSN switches and 2 TSN NICs.}
\label{fig_testbed}
\vspace{-0.3cm}
\end{figure}

We set up a simulated environment on tsnkit \cite{xue2024real}, which is a high-fidelity scheduling and benchmark toolkit for TSN. Simulation is done with an Intel i7-12700KF CPU, 12 cores with a clock speed of 3.60 GHz, and 32 GB DDR4 memory.
We further built a testbed in accordance with IEEE TSN standards as in \S \ref{chap_switch_model} for actual network measurements to better assess the reliability of ARTSN scheduling. It contains 5 switches and 2 NICs, as in Fig. \ref{fig_testbed}. 
The switches and Network Interface Controllers (NICs) are both built on AMD Kintex$^\text{TM}$ UltraScale$^\text{TM}$ FPGA platforms \cite{amdkintex}.

\begin{figure}[t]
    \centering
    \hspace{0.4cm}
    \subfigure[Line topology]{
        \includegraphics[scale=0.11]{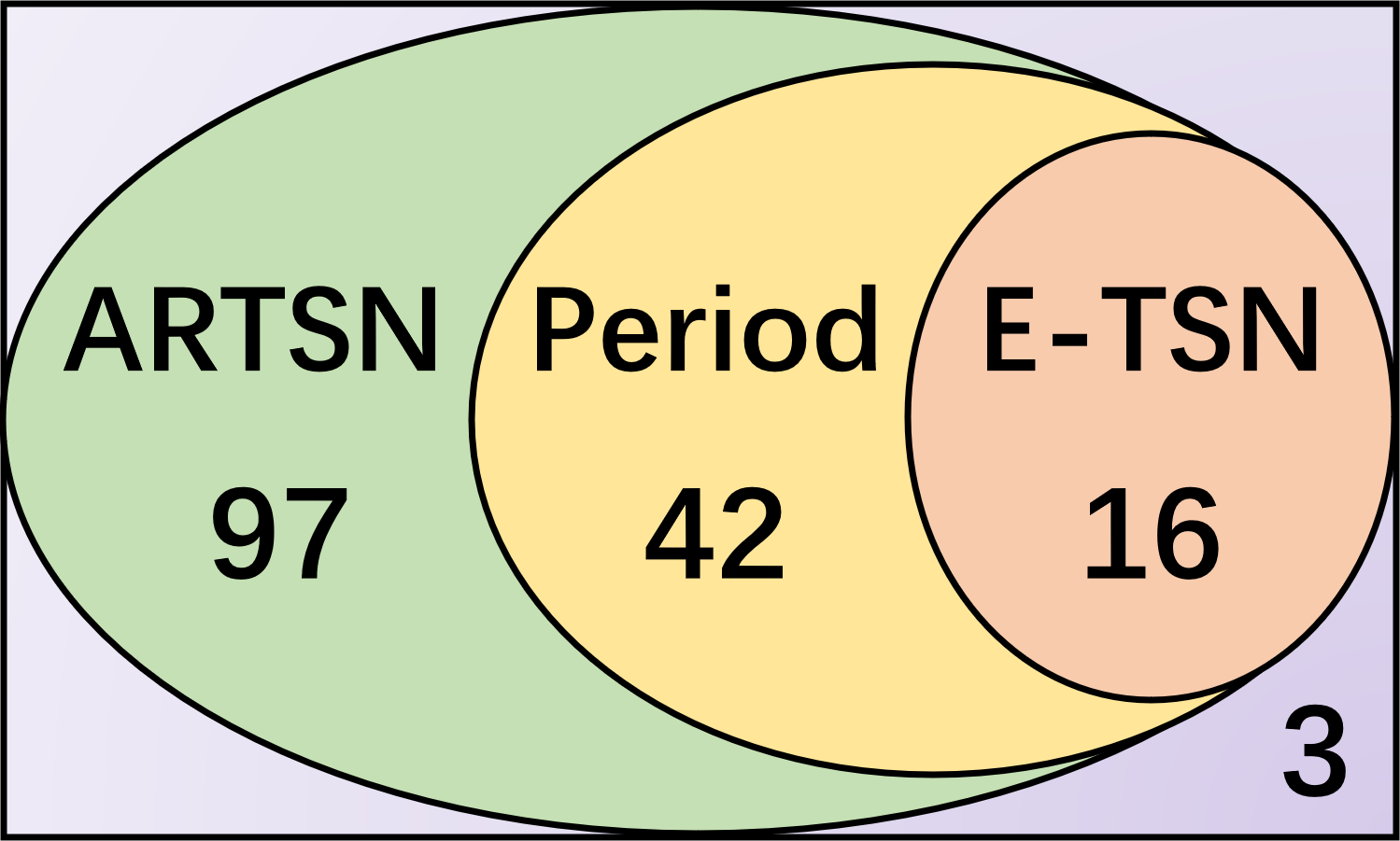}
        \label{fig_sche_line}
    }
    \hspace{0.4cm}
    \subfigure[Star topology]{
        \includegraphics[scale=0.11]{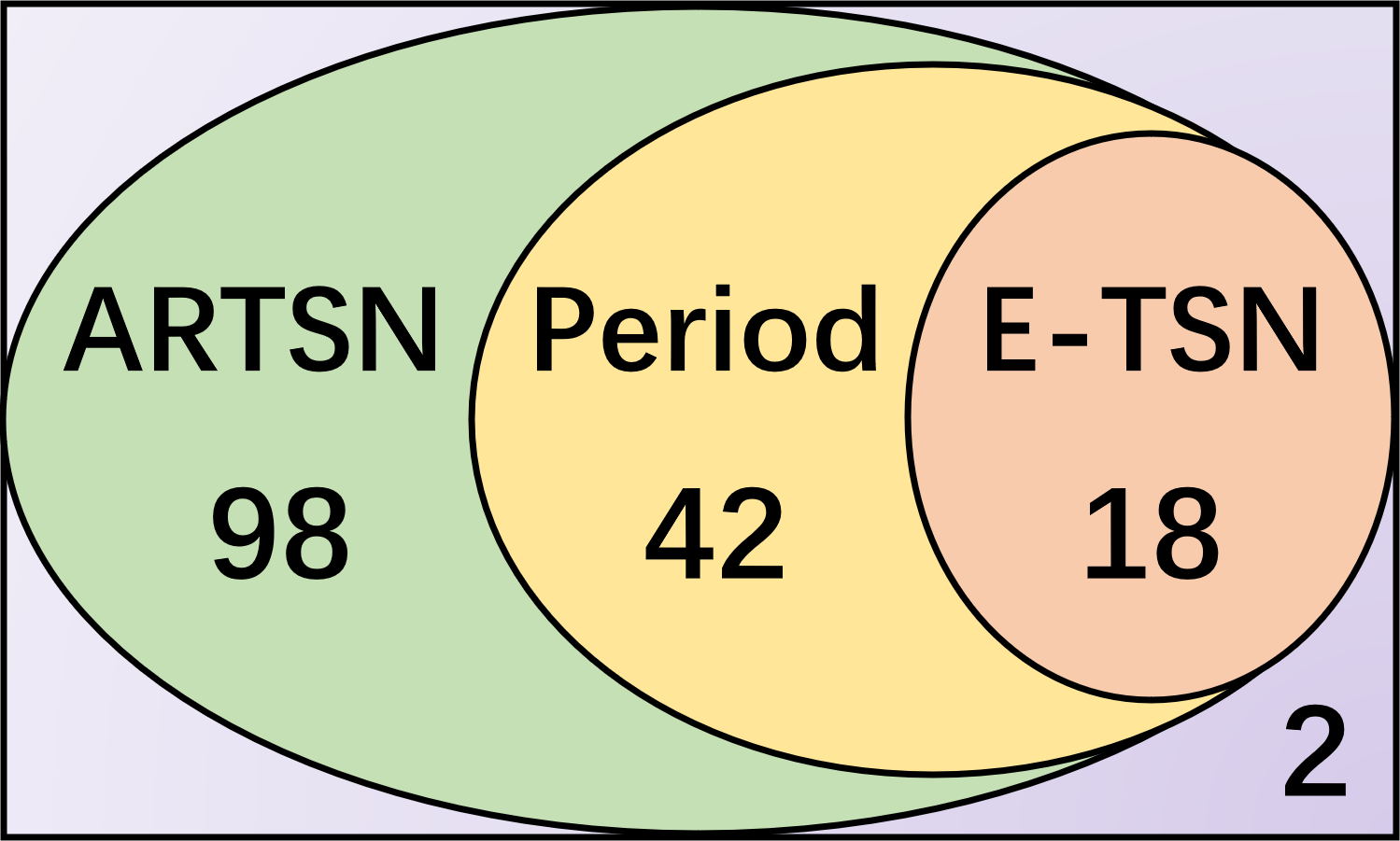}
        \label{fig_sche_star}
    }\caption{Schedulability on the same set of network scenarios.}
    \label{schedulability_result}
\vspace{-0.25cm}
\end{figure} 

\subsection{Baselines}
\label{section_baselines}
We compare ARTSN with the two most relevant state-of-the-art deterministic scheduling methods.
\subsubsection{Period}
Fully deterministic schedules for multi-hop 802.1Qbv-compliant networks are formalized as an SMT problem in \textit{Period} \cite{craciunas2016scheduling}. To compare, ST streams are scheduled as TT streams, with their periods set to 1 ms instead of 10 ms, to achieve lower queueing delay by $10\times$ reservations.
\subsubsection{E-TSN}
Based on \textit{Period}, \textit{E-TSN} \cite{zhao2022etsn} further modeled and addressed ET traffic scheduling. \textit{E-TSN} extends the time slots of each shareable TT stream for the total possible ET traffic that could occur within those slots, and allows ET streams to preempt shareable TT streams, thus guaranteeing their transmission. To compare, ST streams are considered ET during scheduling, \textit{i.e.}, they can arrive at arbitrary moments.

\subsection{Schedulability}
\label{section_schedulability}

We randomly generated 100 network scenarios for each topology, with $U$ ranging from 1 to 50 (two scenarios per $U$), to test the scheduling capabilities of the methods. There is no timeout limit set. For each method, a scenario is unschedulable if there is no feasible scheme that satisfies all its constraints in the entire solution space. A scenario is schedulable as long as one feasible scheme is found.

For the same 100 network scenarios in the line topology, the schedulability of ARTSN, Period, and E-TSN is shown in Fig. \ref{fig_sche_line}. The sets of successfully scheduled scenarios exhibit a containment relationship. The 16 scenarios successfully scheduled by E-TSN are fully contained within the 42 successfully scheduled by Period, which in turn are contained within the 97 successfully scheduled by ARTSN. There are only three scenarios that ARTSN cannot schedule, and they are also not schedulable with other methods. Fig. \ref{fig_sche_star} demonstrates a similar situation in the star topology. E-TSN solves two more scenarios, and ARTSN solves one more. ARTSN’s superior schedulability stems from lower reservation overhead: leveraging deterministic TT–ST triggering, it avoids excessive bandwidth reservations that cause Period and E-TSN to fail under high utilization.

\begin{figure}[t]
    \centering
    \subfigure[Line topology]{
        \includegraphics[scale=0.1232]{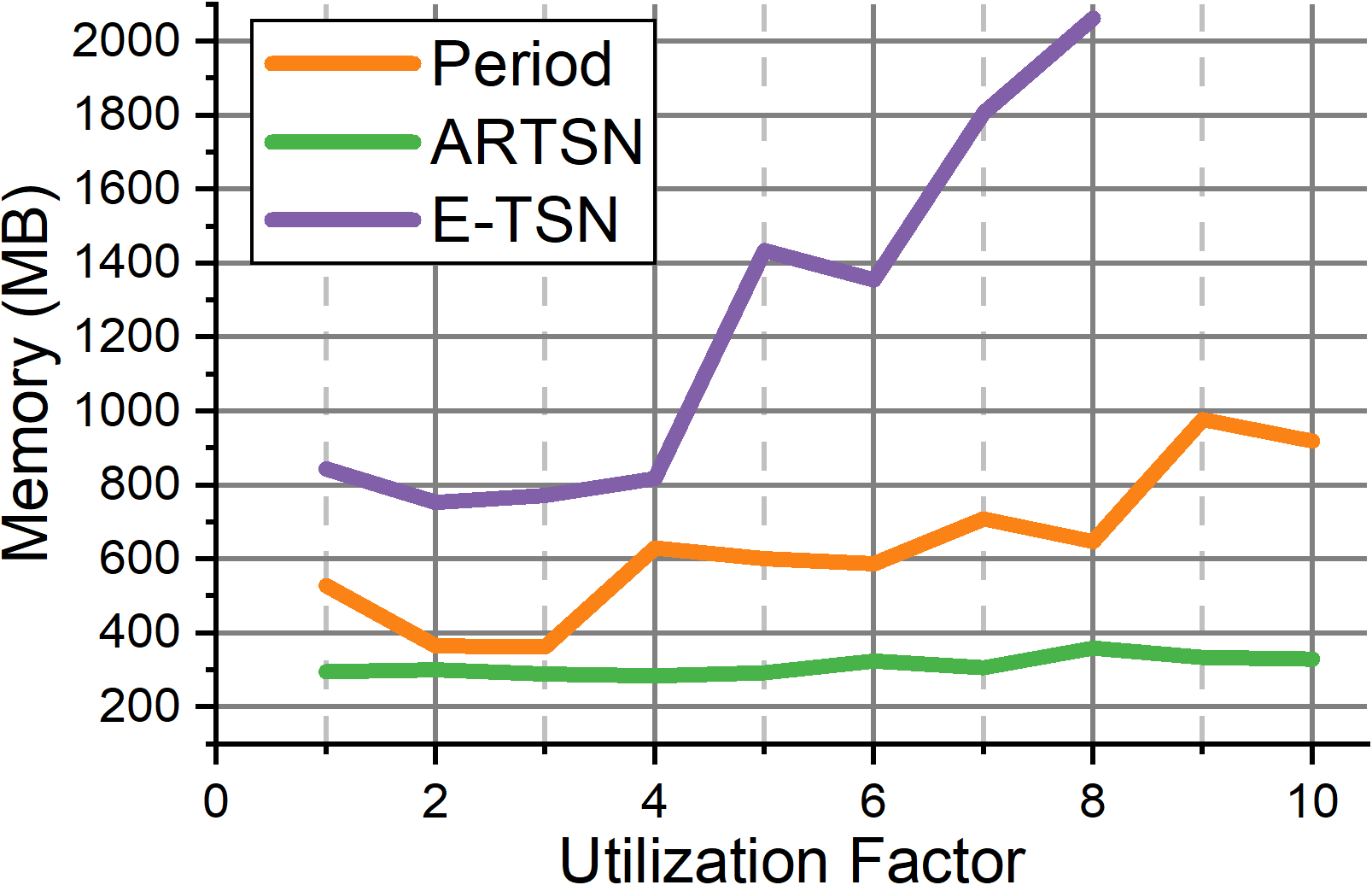}
    }\subfigure[Star topology]{
        \includegraphics[scale=0.1232]{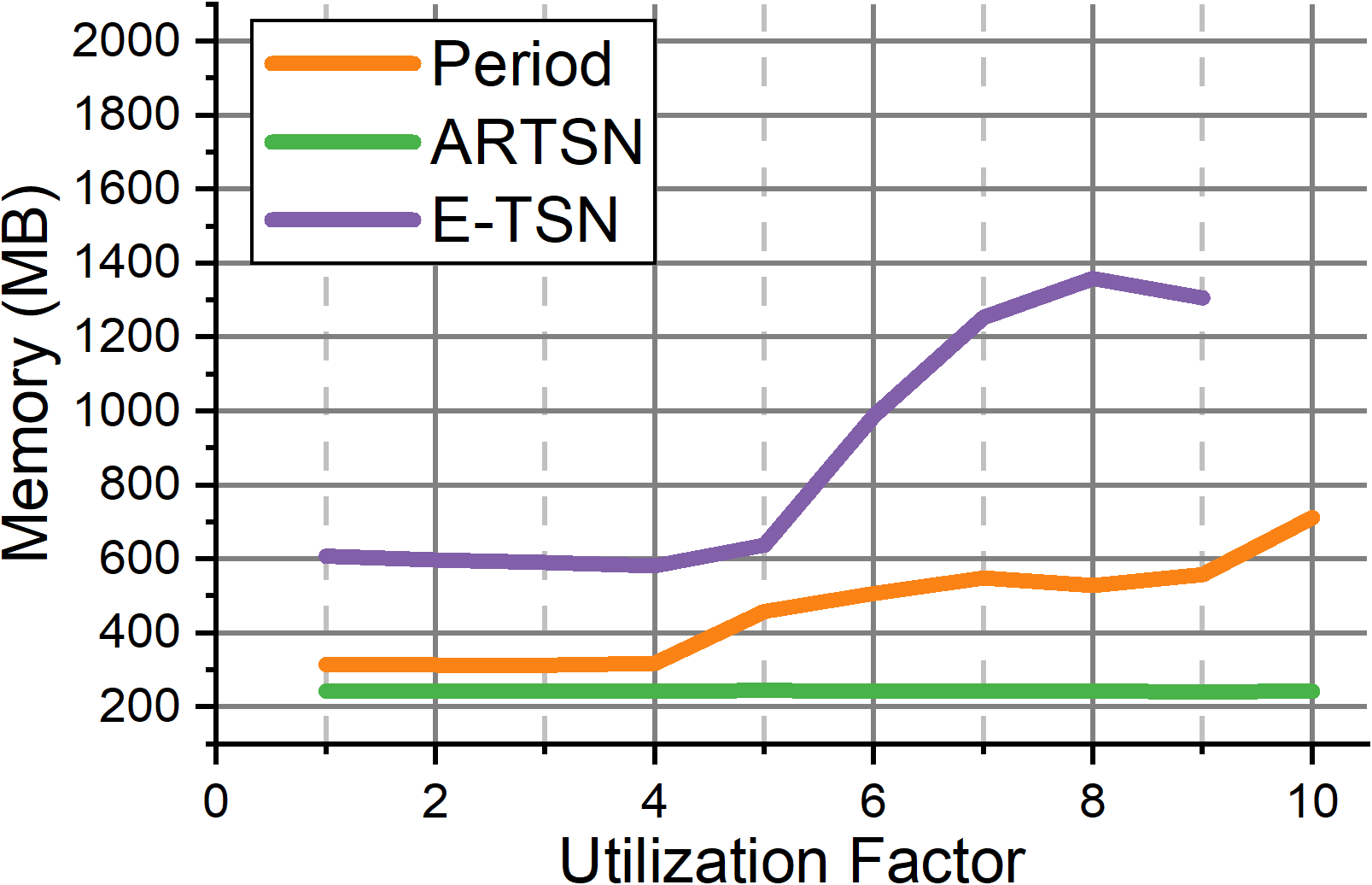}
    }\caption{Memory consumption under different network loads.}
    \label{fig_mem_result}
    \vspace{-0.25cm}
\end{figure}

\subsection{Scalability}
\label{section_scalability}
We compare the scalability of the scheduling methods from both (1) memory consumption and (2) runtime perspectives. As $U$ ranges from 1 to 10, the network is under increasing traffic pressure. The rates at which memory and runtime overhead increase with traffic pressure are key indicators of scalability. Due to limited computing resources, a 4-hour timeout is set, as in \cite{nasri2017exact}.

\begin{figure}[t]
    \centering
    \subfigure[Line topology]{
        \includegraphics[width=0.8\columnwidth]{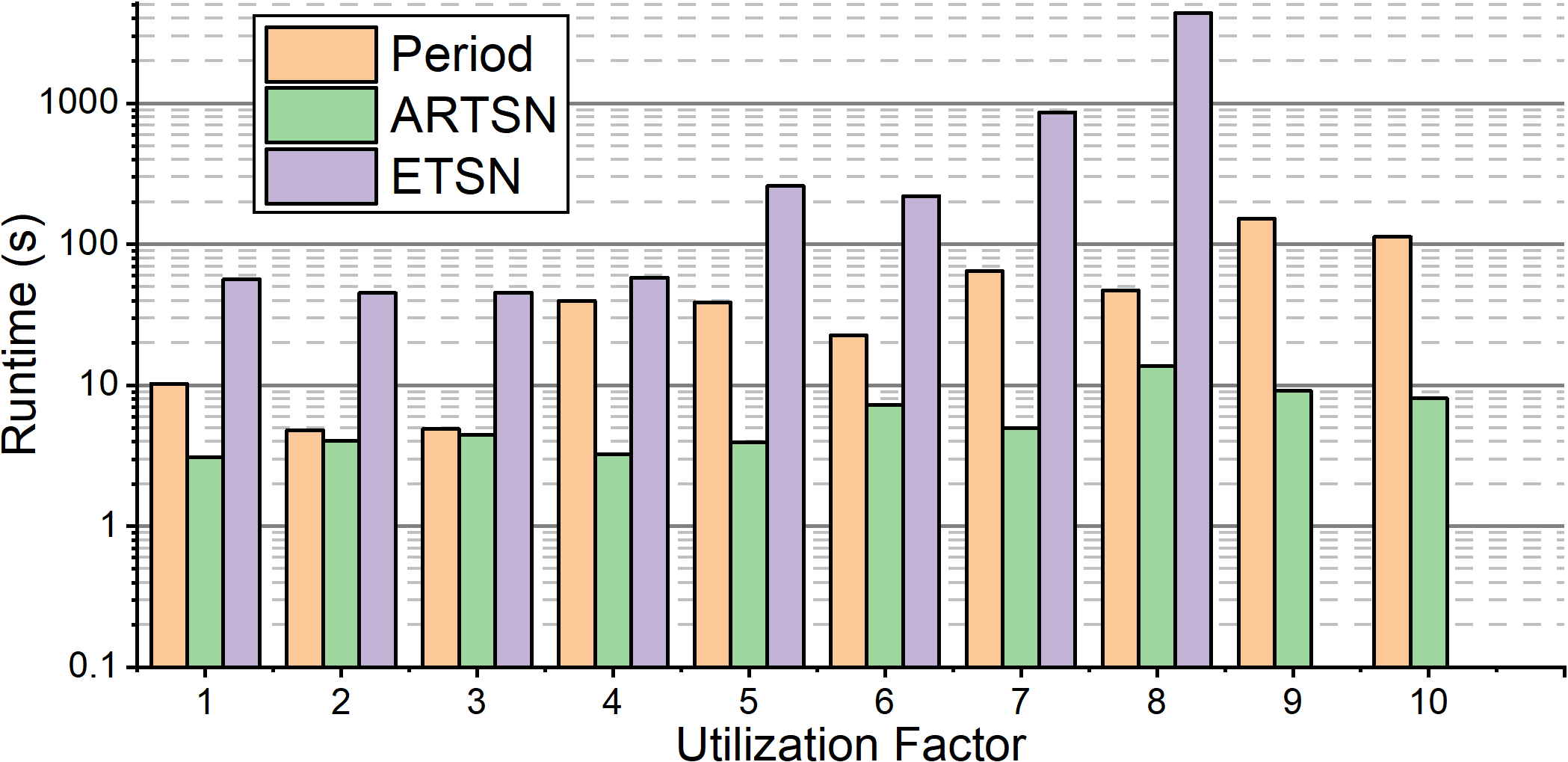}
        \label{fig_line_time}
    }\\
    \vspace{-0.2cm}
    \subfigure[Star topology]{
        \includegraphics[width=0.81\columnwidth]{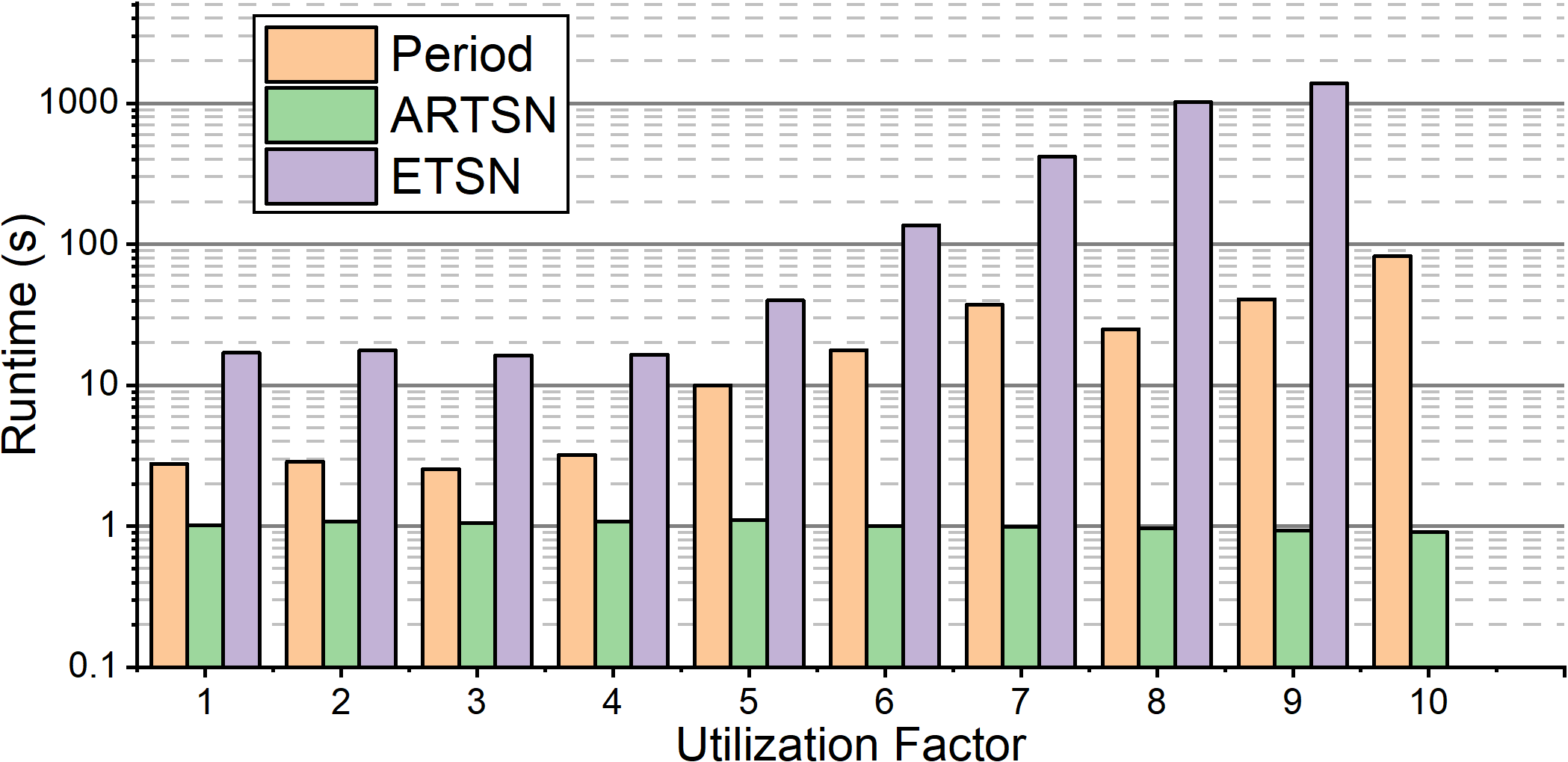}
        \label{fig_star_time}
    }
    \vspace{-0.2cm}
    \caption{Runtime for SMT solving under different network loads. The missing bars in E-TSN are due to scheduling failures or to timeouts.}
    \vspace{-0.3cm}
\end{figure}

The memory consumption of the scheduling methods under different network loads in both line and star topologies is shown in Fig. \ref{fig_mem_result}. In all scenarios, the memory consumption of the three methods shows the same magnitude relationship: E-TSN consumes the most memory, ARTSN the least, and Period is always between them. In the $U=8$ case of line topology, E-TSN consumes 2062.492 MB, Period consumes 647.891 MB, and ARTSN consumes only 359.926 MB. Their memory consumption grows as the network load increases, but at different rates. The consumption of E-TSN shows a nearly linear increase from the $U=4$ case in both topologies, whereas ARTSN is almost unaffected by network load.

Fig. \ref{fig_line_time} and Fig. \ref{fig_star_time} show the runtime of the scheduling methods for the line and star topologies, respectively. In all cases, ARTSN takes at most 13.67 seconds to find a feasible solution, while Period takes quarters, and E-TSN takes hours. In the line topology, the ARTSN runtime increases slightly as network load increases exponentially, whereas it remains almost constant in the star topology. The runtime of Period and E-TSN shows nearly linear growth on a logarithmic scale, \textit{i.e.}, exponential growth on a linear scale. It is worth noting that, even though the network reservation for E-TSN with $U$ of 9 exceeds 100\% (as in Table \ref{table_utilization}), it still schedules successfully in the star topology. This is because E-TSN allows some time slots to overlap each other.

We attribute the dramatically better scalability of ARTSN to the fact that its exact constraints on the ST streams, based on the relationship between the TT streams and the triggered ST streams, significantly reduce the solution space. A smaller search space leads to an easier search and a lower overhead.

\subsection{Efficiency}
\label{section_efficiency}

We define network utilization numerically as
\begin{equation}
Utilization := \max_{\langle n_a, n_b \rangle}\frac{\sum_{s_{i,j}}^{S}\tau_{i,j}^{\langle n_a, n_b \rangle}}{H},
\end{equation}
which represents the proportion of time that the most heavily used link is utilized for transmissions within one hyperperiod. Specifically, it measures how heavily loaded a network is. Table \ref{table_utilization} compares the actual network utilization with the reservation of the three scheduling algorithms. The actual network utilization can be derived from Table \ref{table_flow}.

\begin{table}[tb]
\caption{Utilization and reservation corresponding to $U$ factors}
\centering
\begin{tabular}{cccccc}
\hline
$U$      & 1       & 9        & 20      & 40      & 50        \\ \hline
Actual   & 2.00\%  & 18.00\%  & 40.00\% & 80.00\% & 100.00\%  \\
ARTSN*   & 2.01\%  & 18.01\%  & 40.01\% & 80.01\% & 100.01\%  \\
ARTSN    & 2.08\%  & 18.72\%  & 41.60\% & 83.20\% & 104.00\%  \\
Period   & 4.60\%  & 41.40\%  & 92.00\% & 184.0\% & 230.00\%  \\
E-TSN    & 11.60\% & 104.4\%  & 232.0\% & 464.0\% & 580.00\%  \\ \hline
\end{tabular}
\label{table_utilization}
\end{table}

\begin{figure}[t]
    \centering
    \subfigure[Net throughput gain]{
        \includegraphics[scale=0.21]{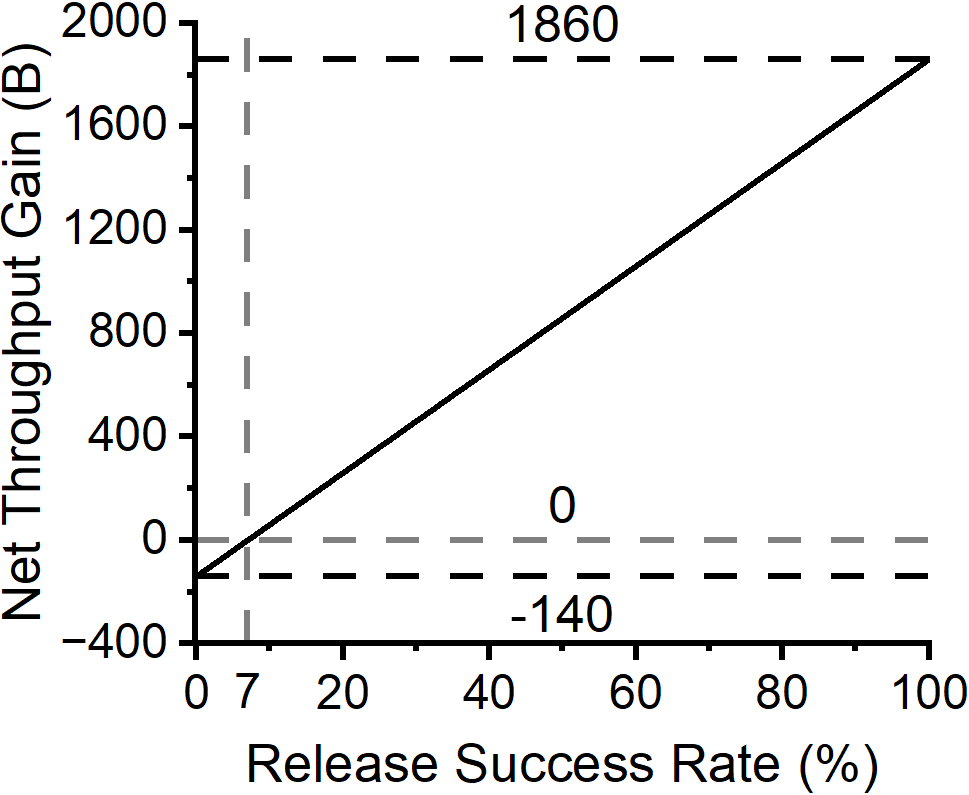}
        \label{fig_net_gain}
    }\subfigure[Max efficiency increase]{
        \includegraphics[scale=0.21]{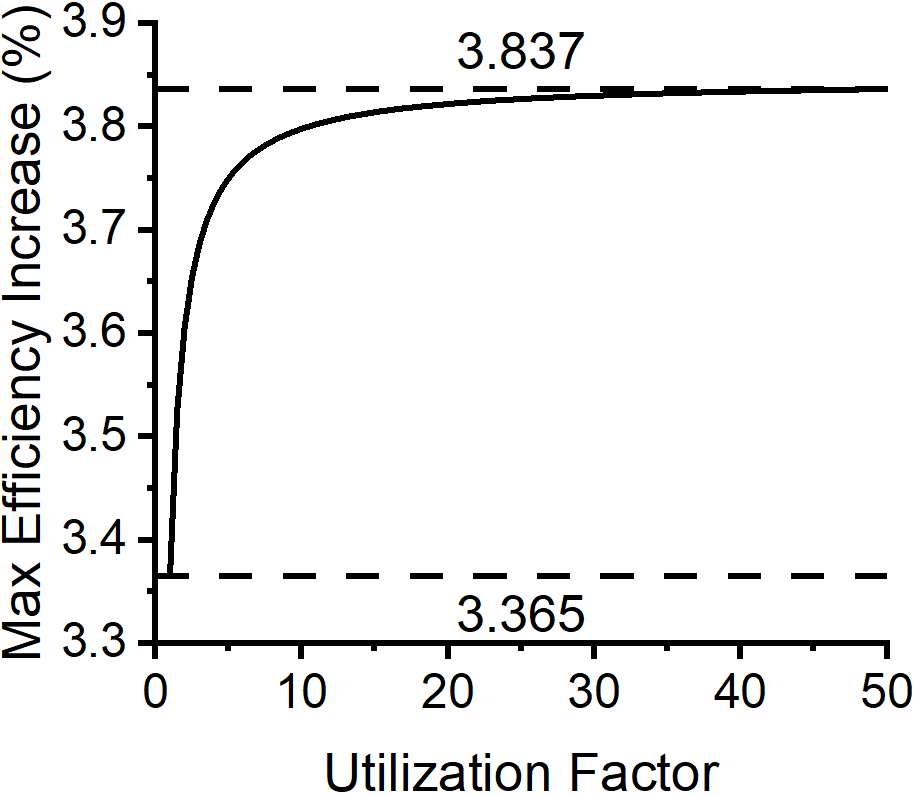}
        \label{fig_max_efficiency}
    }\caption{Results of enabling the adaptive slot release mechanism.}
    \vspace{-0.3cm}
\end{figure} 

\begin{figure*}[tbp]
    \centering
    \subfigure[Line topology, $U$ = 1]{
        \includegraphics[scale=0.144]{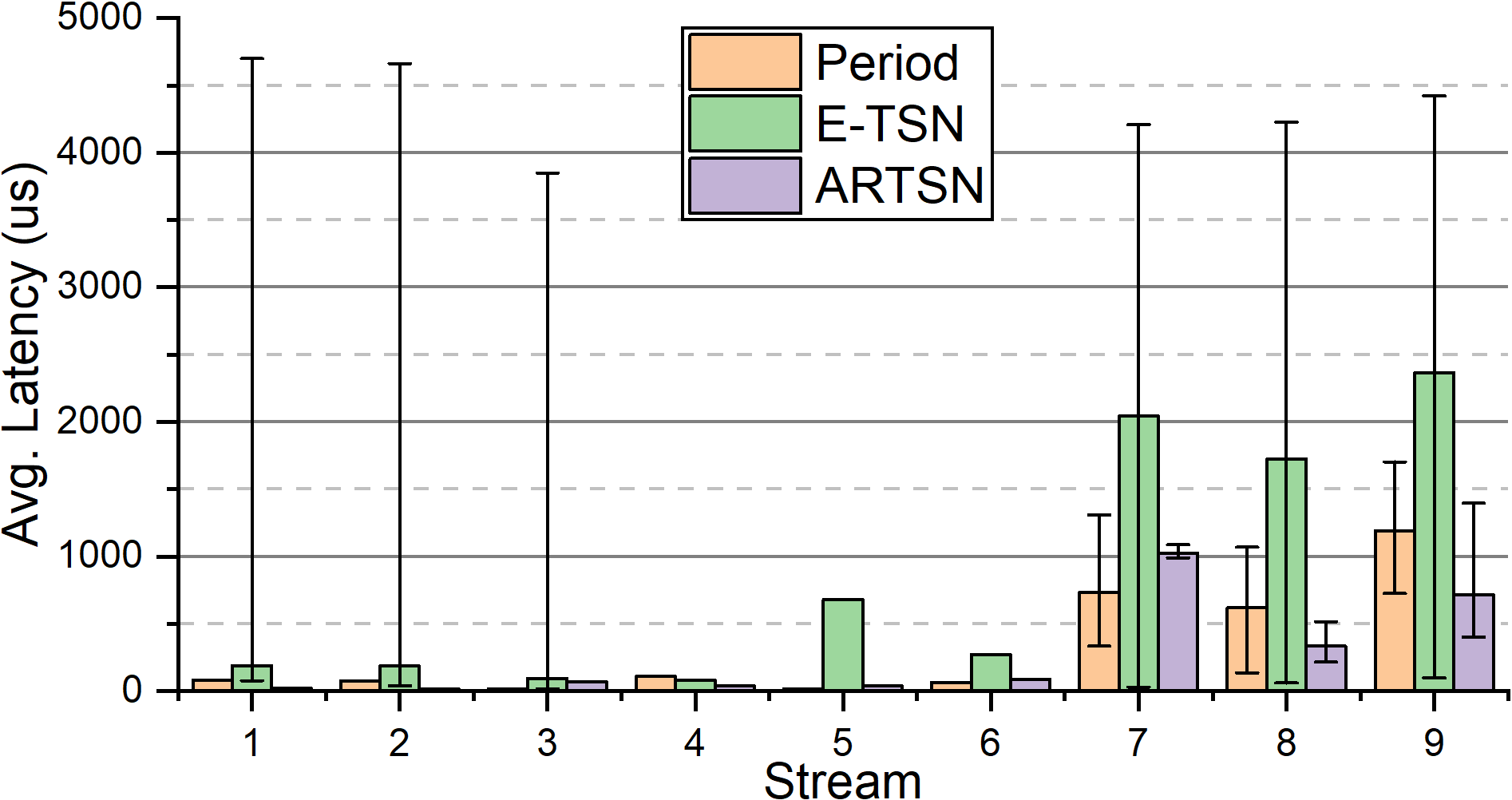}
    }\subfigure[Line topology, $U$ = 8]{
        \includegraphics[scale=0.144]{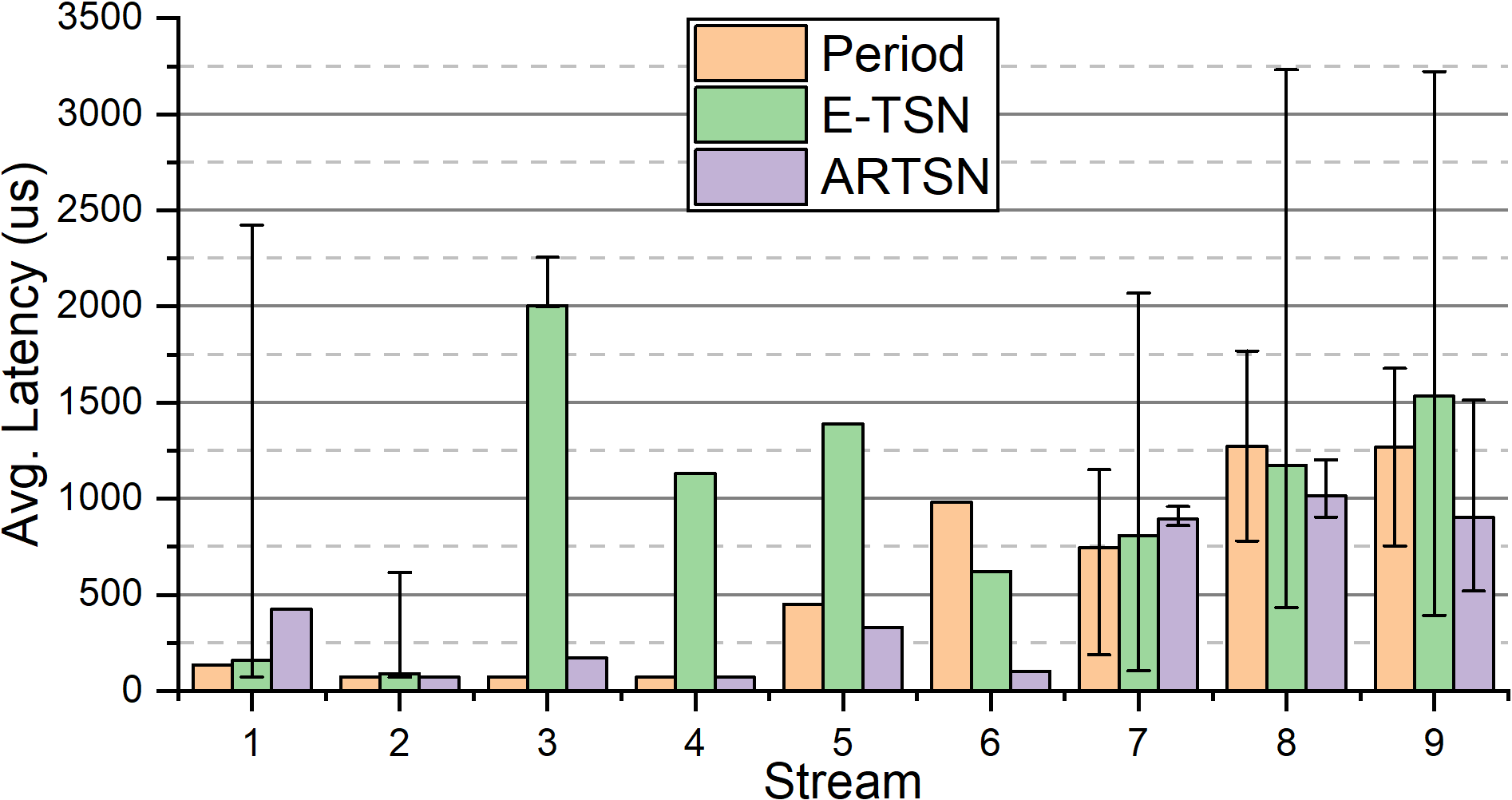}
    }\subfigure[Star topology, $U$ = 8]{
        \includegraphics[scale=0.144]{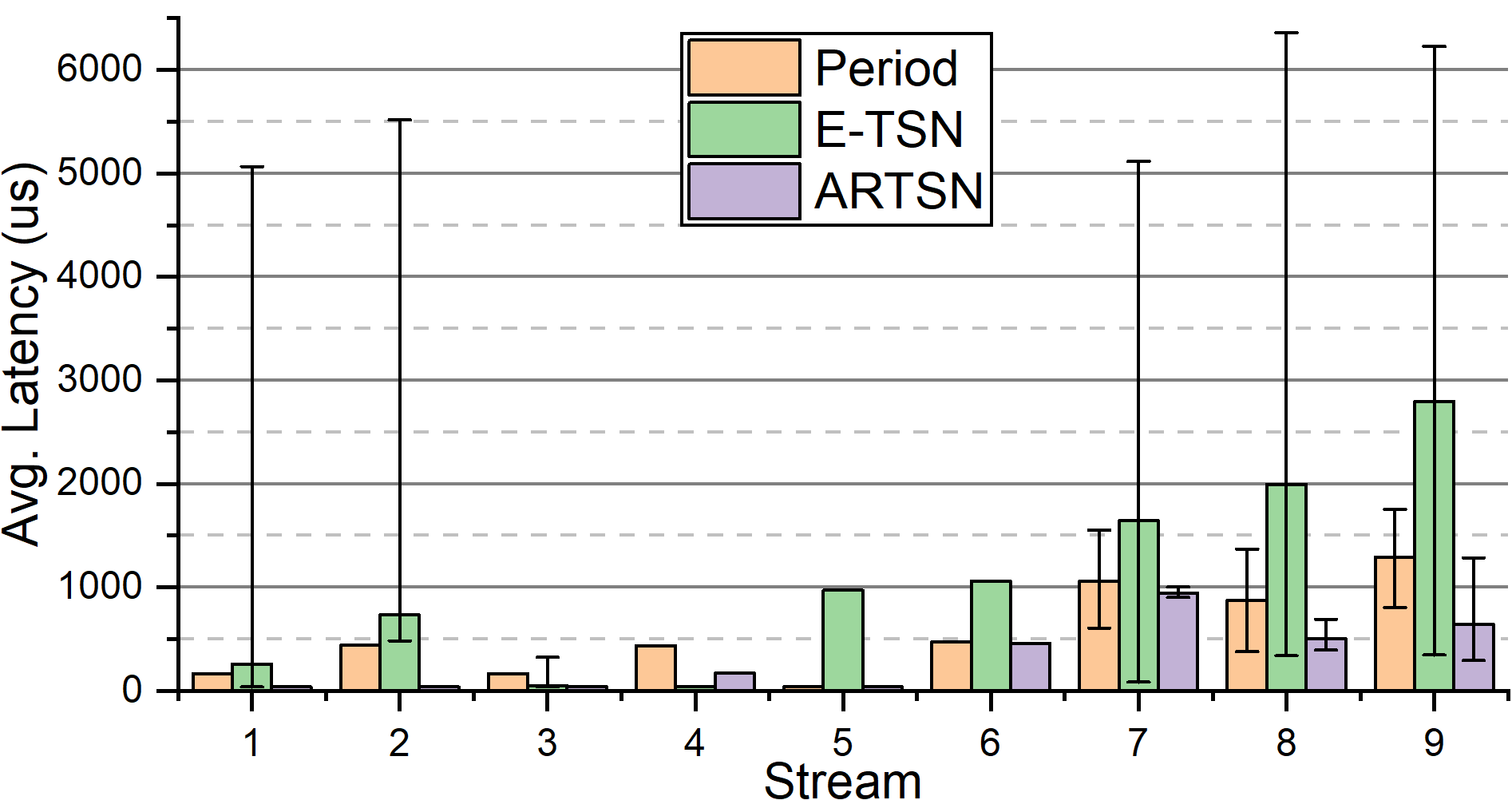}
        \label{fig_U8star}
    }\caption{Average E2E latency of the nine selected streams. Each stream is from one of the different stream classes listed in Table \ref{table_flow}. Streams 1, 2, 3 are shareable TT, streams 4, 5, 6 are strict TT, and streams 7, 8, 9 are ST. The error bars show the range from minimum to maximum latency.}
    \label{fig_stream_result}
\vspace{-0.4cm}
\end{figure*}

In the case of $U=1$, there are 90 TT frames of 500 bytes and 6 ST frames with varying sizes and trigger probabilities in a 20 ms hyperperiod. The expected load is 50 KB per hyperperiod. Given that a 1 Gbps link can transmit 2.5 MB in 20 ms, the actual utilization is 50 KB / 2.5 MB = 2.00\%. With the exact traffic scheduling ability, ARTSN can guarantee this 2.00\% actual traffic at a cost of 2.08\% reservations. So we say ARTSN has an efficiency of $2.00/2.08 = 96.15\%$.

With U up to 40, the actual utilization reaches 80.00\%. ARTSN can complete scheduling at 83.20\%, with only 3.20\% extra reservations. In contrast, Period and E-TSN require 184\% and 464\% reservations, respectively, which cause them to fail in scheduling. Only 43.48\% and 17.24\% of the reserved slots are utilized in Period and E-TSN. The efficiency gap arises because Period treats ST as fully periodic (reserving slots throughout the hyperperiod), and E-TSN treats ST as fully aperiodic (reserving at all possible arrival times). In contrast, ARTSN reserves slots only within the bounded jitter window where ST traffic may arrive, achieving near-optimal efficiency.

Enabling the adaptive slot release mechanism may further reduce unnecessary reservations, as shown by ARTSN* in Table \ref{table_utilization}. When $U = 1$, the expected unused reservation of one hyperperiod is $(50\%*500\text{ B} + 30\%*1500\text{ B} + 20\%*1500\text{ B}) * (20\text{ ms}/10\text{ ms}) = 2000\text{ B}$, while the overhead of sending the RP frames is $(50\% + 30\% + 20\%)*70\text{ B} * (20\text{ ms}/10\text{ ms}) = 140\text{ B}$. RP frames are transmitted as BE traffic. In an ideal situation where every release is successful, ARTSN could achieve a $1860\text{ B}$ net throughput gain for each hyperperiod, as Fig. \ref{fig_net_gain} shows. On the contrary, 140 B is incurred as ineffective overhead if all releases fail. The threshold success rate of a positive throughput gain is $7\%$. Therefore, when the observed success rate exceeds a threshold, ARTSN adaptively activates the slot release mechanism for higher efficiency.

The maximum achievable increase in efficiency depends on the amount of ST traffic and its triggering probabilities. Fig. \ref{fig_max_efficiency} shows how the maximum efficiency increase changes with the $U$ factor varying from 1 to 50. The overhead of sending RP frames is constant, while the releasable time slots scale up. This enables the maximum efficiency increase to grow with the ST traffic load as a concave function, until it converges to $\lim_{U\to\infty}\frac{U*2\text{ KB}-140\text{ B}}{U * 52\text{ KB}}=2/52\approx3.846\%$.

\subsection{Reliability}
\label{section_validation}

To verify its reliability, we used ARTSN to schedule 9 cases, with $U$ ranging from 1 to 40. All scheduling schemes scheduled by ARTSN were verified through both the tsnkit simulator and our 5-hop testbed. In the simulator, each scheme was repeatedly verified over 100 hyperperiods with randomized ST triggering patterns. On the testbed, we first used the NIC-plugged host as the CNC to solve and broadcast GCL, then performed 5-minute E2E observations for each scheme. A worst-case latency of 47.18 µs and a maximum jitter of 1.01 µs were observed, well within the 5 ms deadline constraint. Both results confirm that all types of critical traffic were well protected from 1 Gbps background interference traffic during transmission, with zero deadline misses observed.

We selected 9 streams for a detailed E2E latency comparison, as shown in Fig. \ref{fig_stream_result}. As constrained, no deadlines were missed in any scenario. The latency of strict TT streams was always constant, while the latency of shareable TT streams in E-TSN and ST streams fluctuated within a safe range due to the randomness of preemption and ST triggering.

Benefiting from the preemption of shareable TT streams, E-TSN achieved the minimum possible transmission latency for most ST streams (except for stream 9 in Fig. \ref{fig_U8star}, where ARTSN performs better) but at the cost of a significantly larger jitter. In comparison, the scheduling of Period is more stable, while the scheduling of ARTSN is the most deterministic. In addition, ARTSN schemes always have the lowest upper bounds of latency on ST streams. In terms of reliability, the upper bounds (worst cases) are significantly more important than the average values and the lower bounds.

Compared to the time synchronization error of $< 10$ ns per hop (measured with oscilloscope), the processing delay, transmission delay, and propagation delay of $\sim 10$ µs per hop, the queuing delay can easily amount to $1000+$ µs (inferred from Fig. \ref{fig_stream_result}). The dominance of queuing delay and the decisive role of scheduling schemes are evident. Through exact scheduling schemes, ARTSN guarantees the lowest latency upper bound, thereby ensuring transmission reliability.

\section{Conclusion}
As real-time systems race toward full autonomy, ST control becomes the key enabler—yet it introduces unique challenges to the ARTS network that existing schedulers cannot address. In this work, we present ARTSN to fill the void. It provides a scheduling paradigm that addresses, for the first time, the volatility and absence challenges in handling ST traffic. The integration of exact traffic scheduling and the adaptive slot release mechanism optimizes network resource utilization. Through both simulated and real-world experiments, ARTSN demonstrates significant improvements in schedulability, scalability, and efficiency without compromising reliability. This work bridges a critical gap between autonomous control and deterministic networking, enabling safer and wider ARTS deployments in industrial automation, smart grids, embodied intelligence, and other fields.

\appendix

\subsection{Frame Isolation Constraint}
\label{appendix_frame_iso}
\noindent
\begin{equation}
\begin{aligned}
\forall & \langle n_c,n_d \rangle \in \mathcal{L},\forall s_{i,j}, s_{k,l} \in S, i \ne k,\\
\forall & f_{s_{i,j},u}^{\langle n_c,n_d \rangle} \in F_{s_{i,j}}^{\langle n_c,n_d \rangle},\forall f_{s_{k,l},w}^{\langle n_c,n_d \rangle} \in F_{s_{k,l}}^{\langle n_c,n_d \rangle},\\
\forall & x \in \{1,\dots,lcm(T_i,T_k)/T_i\},\\
\forall & y \in \{1,\dots,lcm(T_i,T_k)/T_k\}:\\
&(\phi_{s_{i,j},u}^{\langle n_c,n_d \rangle} \times \langle n_c,n_d \rangle.tu + x \times T_k \le\\
&\phi_{s_{i,j},u}^{\langle n_a,n_c \rangle} \times \langle n_a,n_c \rangle.tu + \langle n_a,n_c \rangle.pd + y \times T_i) \vee\\
&(\phi_{s_{k,l},w}^{\langle n_c,n_d \rangle} \times \langle n_c,n_d \rangle.tu + y \times T_i \le\\
&\phi_{s_{i,j},u}^{\langle n_b,n_c \rangle} \times \langle n_b,n_c \rangle.tu + \langle n_b,n_c \rangle.pd + x \times T_k) \vee\\
&(q_i \ne q_k)
\end{aligned}
\end{equation}

\subsection{Adjacent Link Constraints}
\label{appendix_adjacent_link}
\noindent
\begin{equation}
\begin{aligned}
\forall s_{i,j} \in & S, \forall \langle n_a,n_b \rangle,\langle n_b,n_c \rangle \in P_i:\\
o & \leftarrow \max (\left | F_{s_{i,j}}^{\langle n_a,n_b \rangle} \right | - \left | F_{s_{i,j}}^{\langle n_b,n_c \rangle} \right |,0)\\
\forall & f_{s_{i,j},k}^{\langle n_b,n_c \rangle} \in F_{s_{i,j}}^{\langle n_b,n_c \rangle}:\\
& \phi_{s_{i,j},k}^{\langle n_b,n_c \rangle} \times \langle n_b,n_c \rangle.tu - \langle n_a,n_b \rangle.pd \ge \\
& (\phi_{s_{i,j},k+o}^{\langle n_a,n_b \rangle} + \tau_{s_{i,j},k+o}^{\langle n_a,n_b \rangle}) \times \langle n_a,n_b \rangle.tu
\\
\\
\end{aligned}
\end{equation}
\vspace{-0.9cm}
\bibliographystyle{IEEEtran}
\bibliography{INFO_ref}

\end{document}